\documentclass[journal]{IEEEtran}

\usepackage[pdftex]{graphicx}
\DeclareGraphicsExtensions{.pdf,.jpeg,.png}

\usepackage[cmex10]{amsmath}
\usepackage{amsfonts}
\usepackage{cite}
\usepackage{url}
\usepackage{soul}
\usepackage{color}
\usepackage{array}
\usepackage{subfigure}
\usepackage{esint} 
\usepackage{tabu}

\newcommand{\BE}{\begin{equation}}   
\newcommand{\EE}{\end{equation}}     
\newcommand{\BEn}{\begin{equation*}} 
\newcommand{\EEn}{\end{equation*}}   
\newcommand{\BF}{\begin{figure}}     
	\newcommand{\EF}{\end{figure}}  	 
\newcommand{\BFw}{\begin{figure*}}   
	\newcommand{\EFw}{\end{figure*}}     
\newcommand{\BT}{\begin{table}}		 
	\newcommand{\ET}{\end{table}}        

\providecommand{\V}[1]{\boldsymbol{#1}}     

\begin{document}
\title{Optimal Currents on Arbitrarily Shaped Surfaces}
\author{Lukas~Jelinek, Miloslav~Capek,~\IEEEmembership{Member,~IEEE}
\thanks{Manuscript received February XX , 2016; revised February XX, 2016.
This work was supported by the Czech Science Foundation
under project \mbox{No.~15-10280Y}.}
\thanks{L.~Jelinek and M.~Capek are with the Department of Electromagnetic Field, Faculty of Electrical Engineering, Czech Technical University in Prague, Technicka 2, 16627, Prague, Czech Republic
(e-mail: lukas.jelinek@fel.cvut.cz).}
}

\markboth{Journal of \LaTeX\ Class Files,~Vol.~XX, No.~XX, February~2016}
{Jelinek \MakeLowercase{\textit{et al.}}: Optimal Currents on Arbitrarily Shaped Surfaces}
\maketitle

\begin{abstract}
An optimization problem has been formulated to find a resonant current extremizing various antenna parameters. The method is presented on, but not limited to, particular cases of gain $G$, quality factor $Q$, gain to quality factor ratio $G/Q$, and radiation efficiency $\eta$ of canonical shapes with conduction losses explicitly included. The Rao-Wilton-Glisson basis representation is used to simplify the underlying algebra while still allowing surface current regions of arbitrary shape to be treated. By switching to another basis generated by a specific eigenvalue problem, it is finally shown that the optimal current can, in principle, be found as a combination of a few eigenmodes. The presented method constitutes a general framework in which the antenna parameters, expressed as bilinear forms, can automatically be extremized.

\end{abstract}

\section{Introduction}
\label{Intro}
\IEEEPARstart{O}{ptimal} radiators, i.e., radiators whose fundamental parameters as gain, fractional bandwidth or radiation efficiency \cite{Balanis_Wiley_2005} are as good as allowed by the laws of physics, have been of interest to applied electromagnetism theorists for almost a century \cite{Oseen1922}. The optimal radiator design is commonly divided into two steps, the first of which sets the optimal current occupying a given spatial region, and the second tries to find realistic support for such a current. Presently, the second step can only be solved via heuristic optimization  \cite{RahmatMichielssen_ElectromagneticOptimizationByGenetirAlgorithms,KoulouridisPsychoudakisVolakis2007,ErentokSigmund2011,RahmatSamii_Kovitz_Rajagopalan-NatureInspiredOptimizationTechniques,CismasuGustafsson_FBWbySimpleFreuqSimulation,HassanWadbroBerggren_TopologyOptimizationOfMetallicAntennas,2014_Liu_SMO} . Three important achievements have, however, been reached in the search for optimal current. 

The first is the realization that a current can always be formed to produce infinite directivity. This observation dates back to \cite{Oseen1922} where it is shown that a current properly distributed on an arbitrarily small region can give rise to an arbitrarily narrow radiation pattern. This result has been rediscovered several times \cite{Schelkunoff_linear_arrays_1943,LaPaz_Superdirectivity_1943,Bouwkamp_superdirectivity_1946} and was given the name superdirectivity. Interest in this topic continues even nowadays \cite{Margetis_superdirectivity_1998,Shamonina_superdirectivity_2015} and reviews can be found in \cite{Bloch_superdirectivity_1953,Hansen_ESA_SuperDirAndSuperconducAntennas}.

The second finding on the optimal current concerns the upper bound of the gain to quality factor ratio $G/Q$. This ratio was proposed as a trade-off between superdirectivity and available bandwidth and was found to be, at a given electrical size, strictly limited \cite{Chu_PhysicalLimitationsOfOmniDirectAntennas} from above. The original work \cite{Chu_PhysicalLimitationsOfOmniDirectAntennas} was latter refined \cite{Fante_G_Q_1992_TAP,Geyi_PhysicalLimitationOfAntenna} but it still suffered from the general issue of having been derived for the ideal case of current distributed within a spherical region. Such an upper bound was, thus, unattainable by common antenna designs. A breakthrough came with the works \cite{GustafssonSohlKristensson_PhysicalLimitationsOfAntennasOfArbitraryShape_RoyalSoc,GustafssonSohlKristensson_IllustrationsOfNewPhysicalBoundOnLinearlyPolAntennas,YaghjianStuart_LowerBoundOnTheQofElectricallySmallDipoleAntennas} showing that it is the dipolar polarisability \cite{Collin_FieldTheoryOfGuidedWaves} of the antenna body that guides the values of the $G/Q$ ratio, regardless of the shape. The upper bound of the $G/Q$ ratio could, thereby, be given for electrically small antennas of arbitrary shape \cite{GustafssonSohlKristensson_PhysicalLimitationsOfAntennasOfArbitraryShape_RoyalSoc,GustafssonSohlKristensson_IllustrationsOfNewPhysicalBoundOnLinearlyPolAntennas,YaghjianStuart_LowerBoundOnTheQofElectricallySmallDipoleAntennas}. The ultimate solution to this optimization problem was, however, known only after \cite{GustafssonCismasuJonsson_PhysicalBoundsAndOptimalCurrentsOnAntennas_TAP,Gustafsson_OptimalAntennaCurrentsForQsuperdirectivityAndRP} where it was shown that the upper bound of the $G/Q$ ratio can be formulated as a convex optimization problem \cite{BoydVandenberghe_ConvexOptimization}. The versatility of this optimization scheme is presented in \cite{GustafssonTayliEhrenborgEtAl_AntennaCurrentOptimizationUsingMatlabAndCVX}. A similar method is also used in \cite{Geyi_OptimizationOfTheRationOfGainToQ} for wire radiators.

The third important field of work lay in search for current minimizing quality factor $Q$. The pioneering work \cite{Chu_PhysicalLimitationsOfOmniDirectAntennas} set the absolute lower bound of the radiation quality factor $Q$ for current occupying a spherical region of a given electrical size. The approximations used in \cite{Chu_PhysicalLimitationsOfOmniDirectAntennas} were later refined in \cite{McLean_AReExaminationOfTheFundamentalLimitsOnTheRadiationQofESA}, but as with the other optimized antenna parameters, the final goal was to specify the lower bound of quality factor $Q$ for regions of arbitrary shape. Related work began in \cite{YaghjianStuart_LowerBoundOnTheQofElectricallySmallDipoleAntennas}
where the problem was converted to the optimization of the dipole moment of the antenna body when illuminated by a uniform field. The validity of such an approach is restricted to electrically small antennas and suffers from the lack of the method for finding the optimal current. Similar line of reasoning, but using scattering cross-section, was later used in \cite{Thal_QboundsForArbitrarySmallAntennas}. Progress was made by \cite{Vandenbosch_SimpleProcedureToDeriveLowerBoundsForRadiationQofESDofArbitraryTopology} by formulating an explicit condition on the optimal current in electrically small regions of arbitrary shape, even though a general method for obtaining the optimal current was not presented. Recent works \cite{JonssonGustafsson_StoredEnergiesInElectricAndMagneticCurrentDensities_RoyA,Kim_LowerBoundsOnQForFinizeSizeAntennasOfArbitraryShape,ChalasSertelVolakis2016EarlyAccess} approached the lower bound of the radiation quality factor $Q$ on surfaces of arbitrary shape, but a general prescription for the optimal current is still missing.

In this paper an optimization problem for finding resonant current extremizing various antenna parameters is formulated and its solution is presented on the case of quality factor $Q$, gain $G$, gain to quality factor ratio $G/Q$ and radiation efficiency $\eta$. The solution to the problem is given in the Rao-Wilton-Glisson (RWG) basis representation \cite{RaoWiltonGlisson_ElectromagneticScatteringBySurfacesOfArbitraryShape} which allows for describing surfaces of arbitrary shape.

The paper is organized as follows. Section~\ref{math} introduces the necessary mathematical tools. Particularly, Section~\ref{sources} introduces a description by source current, Section~\ref{matrices} introduces a representation of smooth operators by matrices, and Section~\ref{optim} contains a general formulation of the optimization problem to be solved. Section~\ref{results} presents a solution to various optimization tasks covering convex as well as non-convex antenna parameters. Section~\ref{excit} addresses feeding of the optimal current density. The achieved results are discussed in Section~\ref{disc}. The paper concludes in Section~\ref{concl}.

\section{Mathematical Tools}
\label{math}

\subsection{Source Description}
\label{sources}
The description of an electromagnetic system can be equivalently given in terms of fields or field sources \cite{Schwinger_ClassicalElectrodynamics,Schwinger_Particles_Sources_and_Fields}. As an example, we mention the field formulation of the cycle mean radiated power in a time-harmonic steady state \cite{Harrington_TimeHarmonicElmagField}
\BE
\label{eqS11}
{P_{{\mathrm{rad}}}} = \frac{1}{2}\mathrm{Re} \left\{ {\oint\limits_S {\left( {{\V{E}} \times {{\V{H}}^*}} \right) \cdot {\mathrm{d}}{\V{S}}} } \right\},
\EE
where $\V{E}$, $\V{H}$ are the electric and magnetic intensity within the time-harmonic convention \mbox{$\mathcal{F} \left( t \right)=\mathrm{Re} \left\{F \left( \omega \right) \mathrm{exp} \left(\mathrm{j} \omega t \right)\right\} $} with angular frequency $\omega$. This convention is also used throughout the paper.

Being equivalent to (\ref{eqS11}), the radiated power can also be expressed in terms of field sources as \cite{Schwinger_ClassicalElectrodynamics}
\BE
\label{eqS13}
{P_{{\mathrm{rad}}}} = -\frac{{\mathrm{1}}}{{8{\mathrm{\pi }}{\varepsilon _0}\omega }} \mathrm{Im} \Big \{ {{k^2}\left\langle {{\boldsymbol{J}},L{\boldsymbol{J}}} \right\rangle  - \left\langle {\nabla  \cdot {\boldsymbol{J}},L\nabla  \cdot {\boldsymbol{J}}} \right\rangle } \Big \},
\EE
where $k$ is the wavenumber and a homogeneous, isotropic medium of permittivity $\epsilon$ and permeability $\mu$ is assumed. The prescription (\ref{eqS13}) also involves the operator
\BE
	\label{eqS14a}
	L{\boldsymbol{J}} = \int\limits_{V'} {{\boldsymbol{J}}\left( {{\boldsymbol{r'}}} \right)\frac{{{{\mathrm{e}}^{ - {\mathrm{j}}k\left| {{\boldsymbol{r}} - {\boldsymbol{r'}}} \right|}}}}{{\left| {{\boldsymbol{r}} - {\boldsymbol{r'}}} \right|}}{\mathrm{d}}V'}
\EE
and a suitably defined scalar product
\BE
\begin{aligned}
	\label{eqS12}
	\left\langle {{\boldsymbol{M}},{\boldsymbol{N}}} \right\rangle  = \int\limits_V {{\boldsymbol{M}}^*\left( {\boldsymbol{r}} \right) \cdot {{\boldsymbol{N}}}\left( {\boldsymbol{r}} \right){\mathrm{d}}V}.
\end{aligned}
\EE

The formulation (\ref{eqS13}) affords many advantages, the most important being the change of the integration domain to the source region only.  It is also noteworthy to mention that, at the modest expense of introducing magnetic current \cite{Balanis1989}, the formulation (\ref{eqS13}) can also be rewritten solely in terms of sources distributed on surfaces (thanks to the surface equivalence principle \cite{Balanis1989}) which significantly reduces evaluation time. Due to these properties, the source formulation is a popular choice for optimizations \cite{GustafssonSohlKristensson_PhysicalLimitationsOfAntennasOfArbitraryShape_RoyalSoc,
GustafssonSohlKristensson_IllustrationsOfNewPhysicalBoundOnLinearlyPolAntennas,
Vandenbosch_SimpleProcedureToDeriveLowerBoundsForRadiationQofESDofArbitraryTopology,
YaghjianStuart_LowerBoundOnTheQofElectricallySmallDipoleAntennas,
GustafssonCismasuJonsson_PhysicalBoundsAndOptimalCurrentsOnAntennas_TAP,
Gustafsson_OptimalAntennaCurrentsForQsuperdirectivityAndRP,
GustafssonFridenColombi_AntennaCurrentOptimizationForLossyMediaAWPL,
Gustaffson_QdisperssiveMedia_arXiv,
JonssonGustafsson_StoredEnergiesInElectricAndMagneticCurrentDensities_RoyA,
Kim_LowerBoundsOnQForFinizeSizeAntennasOfArbitraryShape}
and this paper is no exception.  This text therefore assumes \mbox{${{\boldsymbol{J}}{\mathrm{d}}V} \to {{\boldsymbol{J}}{\mathrm{d}}S}$}, switching the units of current density as \mbox{$\left[\mathrm{Am}^{-2}\right] \to \left[\mathrm{Am}^{-1}\right]$}, i.e., to electric sources distributed on a surface.

\subsection{Matrix Description}
\label{matrices}

To reduce the computational burden further, it is advantageous to discretize the continuous quantities presented in Section~\ref{sources}, i.e., to represent them in a (reduced) basis \cite{Cohen-TannoudjiDiuLaloe1992}. In this paradigm, the continuous bilinear forms are transformed as \mbox{$\left\langle {{\boldsymbol{J}},L{\boldsymbol{J}}} \right\rangle  \to {{\mathbf{I}}^{\mathrm{H}}}{\mathbf{LI}}$}, where ${\mathbf{I}}$ is the vector representation of the current density, ${\mathbf{L}}$ is the matrix representation of the operator $L$ and super-index $^{\mathrm{H}}$ represents the Hermitian conjugation. Within the community of applied electromagnetism, such a matrix representation can be dated back to the works of Harrington \cite{Harrington_MatrixMethodsForFieldProblems} and is, once again, becoming increasingly popular \cite{Gustafsson_OptimalAntennaCurrentsForQsuperdirectivityAndRP,Gustaffson_QdisperssiveMedia_arXiv,GustafssonFridenColombi_AntennaCurrentOptimizationForLossyMediaAWPL,Kim_LowerBoundsOnQForFinizeSizeAntennasOfArbitraryShape}.

Throughout this paper, the RWG representation \cite{RaoWiltonGlisson_ElectromagneticScatteringBySurfacesOfArbitraryShape} is used as a primary, in which the surfaces are decomposed into triangular patches and expansion coefficients ${\mathbf{I}}$ are the values of the RWG edge surface current densities.

Notable examples of the RWG representations of continuous bilinear forms used in this paper are:
\begin{itemize}
	\item Radiated power and reactive power \cite{Harrington_TimeHarmonicElmagField}
\end{itemize}
\BE
	\label{eqS21a}
	P_{\mathrm{rad}}+\mathrm{j}P_{\mathrm{react}} \to \frac{1}{2}{{\mathbf{I}}^{\mathrm{H}}}{\left(\mathbf{R}+\mathrm{j}\mathbf{X}\right)\mathbf{I}}.
\EE
\begin{itemize}
	\item Stored electromagnetic energy \cite{Vandenbosch_ReactiveEnergiesImpedanceAndQFactorOfRadiatingStructures}
\end{itemize}
\BE
\label{eqS22}
	{W_{{\mathrm{sto}}}} \to {{\mathbf{I}}^{\mathrm{H}}}{\mathbf{W I}} = {{\mathbf{I}}^{\mathrm{H}}}\frac{1}{4}\frac{{\partial {\mathbf{X}}}}{{\partial \omega }}{\mathbf{I}}.
\EE
\begin{itemize}
	\item Ohmic lost power \cite{Harrington_TimeHarmonicElmagField}
\end{itemize}
\BE
\label{eqS23}
	P_{\mathrm{lost}} \to \frac{1}{2}{{\mathbf{I}}^{\mathrm{H}}}{\mathbf{\Sigma I}}.
\EE
\begin{itemize}
	\item Partial radiation intensity in the direction $\boldsymbol{\theta}_0$ or $\boldsymbol{\varphi}_0$ \cite{Balanis_Wiley_2005}
\end{itemize}
\BE
	\label{eqS24}
	{U_{\left( {\theta /\varphi } \right)}} \to {{\mathbf{I}}^{\mathrm{H}}}{\mathbf{U}_{\left( {\theta /\varphi } \right)} \mathbf{I}}.
\EE
The matrices $\mathbf{R}$ and $\mathbf{X}$ coincide with the real and imaginary parts of the EFIE impedance matrix \cite{Harrington_FieldComputationByMoM}. The construction of matrices $\mathbf{R}$, $\mathbf{X}$, $\mathbf{W}$, $\mathbf{\Sigma}$, $\mathbf{U}_{\left( {\theta /\varphi} \right)}$ can be determined from the smooth forms of the corresponding physical quantities. A detailed description is provided in Section~\ref{app} and can be ignored without going off track in subsequent developments. One only has to keep in mind the physical meaning of bilinear forms (\ref{eqS21a}--\ref{eqS24}).

Using the above description, important radiation characteristics \cite{Balanis_Wiley_2005} can be written as:

\begin{itemize}
	\item Radiation quality factor 
\end{itemize}
\BE
\label{eqS26}
Q = \frac{2\omega{{{\mathbf{I}}^{\mathrm{H}}{\mathbf{W}}{{\mathbf{I}}}} }}{{ {{\mathbf{I}}^{\mathrm{H}}{\mathbf{R}}{{\mathbf{I}}}} }} +
\frac{\left|{{{\mathbf{I}}^{\mathrm{H}}{\mathbf{X}}{{\mathbf{I}}}} }\right|}{2{ {{\mathbf{I}}^{\mathrm{H}}{\mathbf{R}}{{\mathbf{I}}}} }}.
\EE

\begin{itemize}
	\item Partial directivity
\end{itemize}
\BE
\label{eqS27}
D_{\left( {\theta /\varphi } \right)} = \frac{8\pi{{{\mathbf{I}}^{\mathrm{H}}{\mathbf{U}_{\left( {\theta /\varphi } \right)}}{{\mathbf{I}}}} }}{{ {{\mathbf{I}}^{\mathrm{H}}{\mathbf{R}}{{\mathbf{I}}}} }}.
\EE

\begin{itemize}
	\item Radiation efficiency
\end{itemize}
\BE
\label{eqS27}
\eta = \frac{{{{\mathbf{I}}^{\mathrm{H}}{\mathbf{R}}{{\mathbf{I}}}} }}{{ {{\mathbf{I}}^{\mathrm{H}}{\left(\mathbf{R}+\mathbf{\Sigma}\right)}{{\mathbf{I}}}} }}.
\EE
\subsection{Optimization Procedure}
\label{optim}

The description of the antenna characteristics by bilinear forms developed in the previous section gives us the possibility to perform various optimization tasks. A solution to quite a general class of antenna optimization problems (which include the maximization of the gain to quality factor ratio, the minimization of quality factor, the maximization of gain or the maximization of radiation efficiency) is presented in this subsection.

Considering this point, assume there are three Hermitian matrices $\mathbf{A}$, $\mathbf{B}$, $\mathbf{C}$ and that the following optimization problem needs to be solved
\begin{subequations}
	\begin{align}
	\label{eqS31a}
	\displaystyle \mathop {\min}\limits_{\mathbf{I}} \left\{ {{\mathbf{I}}^{\mathrm{H}}}{\mathbf{A I}} \right\},	\\
	\label{eqS31b}
	{{\mathbf{I}}^{\mathrm{H}}}{\mathbf{B I}} =1,\\
	\label{eqS31c}
	{{\mathbf{I}}^{\mathrm{H}}}{\mathbf{C I}} =\gamma,
	\end{align}
\end{subequations}
where column vector ${\mathbf{I}}$ stands for the unknown vector of RWG edge currents, while matrices $\mathbf{A}$, $\mathbf{B}$, $\mathbf{C}$ can be any of the matrices from Section~\ref{matrices}.

At this point it is advantageous to make a transformation to yet another basis, namely to express
\BE
\label{eq32}
{\mathbf{I}} = \sum\limits_n {{\alpha _n}{{\mathbf{I}}_n}}
\EE
with
\BE
\label{eq33}
{\mathbf{S}}{{\mathbf{I}}_n} = {\zeta _n}{\mathbf{T}}{{\mathbf{I}}_n}, 
\EE
where $\mathbf{S}$ and $\mathbf{T}$ are also Hermitian matrices.
The hermicity of matrices assures \cite{Wilkinson_AlgebraicEigenvalueProblem} that vectors ${\mathbf{I}}_n$ form a basis, which can be orthogonalized as
\BE
\label{eqS34a}
{\mathbf{I}}_n^{\mathrm{H}}{\left(\mathbf{T}+\mathrm{j}\mathbf{S}\right)}{{\mathbf{I}}_m} = \left(1+\mathrm{j}\zeta _n\right) {\delta _{mn}}
\EE
and that the eigen-values $\zeta _n$ are real.
The representation (\ref{eq32}) recasts the optimization problem (\ref{eqS31a})--(\ref{eqS31c}) to
\begin{subequations}
\begin{align}
\label{eqS36a}
\displaystyle \mathop {\min}\limits_{\boldsymbol{\alpha}} \left\{ {{\boldsymbol{\alpha}}^{\mathrm{H}}}{\mathbf{A}^{\mathrm{GEP}} \boldsymbol{\alpha}} \right\},	\\
\label{eqS36b}
{{\boldsymbol{\alpha }}^{\mathrm{H}}} \mathbf{B}^{\mathrm{GEP}}  {\boldsymbol{\alpha }} = 1, \\
\label{eqS36c}
{{\boldsymbol{\alpha }}^{\mathrm{H}}}\mathbf{C}^{\mathrm{GEP}}{\boldsymbol{\alpha }} = \gamma, 
\end{align}
\end{subequations}
with
\BE
\label{eqS37}
{{\mathbf{A}}^{{\mathrm{GEP}}}} = \left[ {\begin{array}{*{20}{c}}
	{{\mathbf{I}}_1^{\mathrm{H}}{\mathbf{A}}{{\mathbf{I}}_1}}& \ldots &{{\mathbf{I}}_1^{\mathrm{H}}{\mathbf{A}}{{\mathbf{I}}_N}}\\
	\vdots & \ddots & \vdots \\
	{{\mathbf{I}}_N^{\mathrm{H}}{\mathbf{A}}{{\mathbf{I}}_1}}& \cdots &{{\mathbf{I}}_N^{\mathrm{H}}{\mathbf{A}}{{\mathbf{I}}_N}}
	\end{array}} \right]
\EE
as the representation of matrix $\mathbf{A}$ in the basis (\ref{eq33}), the superscript $^{{\mathrm{GEP}}}$ denoting a generalized eigenvalue problem and with column vector ${\boldsymbol{\alpha}}$ as the representation of vector $\mathbf{I}$. The meaning of ${{\mathbf{B}}^{{\mathrm{GEP}}}}$, ${{\mathbf{C}}^{{\mathrm{GEP}}}}$ is analogous to ${{\mathbf{A}}^{{\mathrm{GEP}}}}$.

The solution to (\ref{eqS36a})--(\ref{eqS36c}) is found by the method of Lagrange multipliers \cite{NocedalWright_NumericalOptimization}. The corresponding Lagrangian density reads
\BE
\begin{aligned}
\label{eqS38}\
{\cal L}\left( {{\boldsymbol{\alpha }},{\lambda _1},{\lambda _2}} \right) &= {{\boldsymbol{\alpha }}^{\mathrm{H}}}{{\boldsymbol{A}}^{{\mathrm{GEP}}}}{\boldsymbol{\alpha }} - {\lambda _1}\left( {{{\boldsymbol{\alpha }}^{\mathrm{H}}}{{\mathbf{C}}^{{\mathrm{GEP}}}}{\boldsymbol{\alpha }} - \gamma } \right) \\
 &- {\lambda _2}\left( {{{\boldsymbol{\alpha }}^{\mathrm{H}}} {{\mathbf{B}}^{{\mathrm{GEP}}}} {\boldsymbol{\alpha }} - 1} \right).
\end{aligned}
\EE
Note that the hermicity of the matrices $\mathbf{A}$, $\mathbf{B}$, $\mathbf{C}$ assures that \mbox{${\cal L}\left( {{\boldsymbol{\alpha }},{\lambda _1},{\lambda _2}} \right) \in \mathbb{R}$} for \mbox{$\lambda_1,\lambda_2\in \mathbb{R}$}, while vectors $\boldsymbol{\alpha}$ are generally complex. The stationary points of (\ref{eqS38}) follow from
\BE
\label{eqS39}
\frac{{\partial {\cal L}}}{{\partial \alpha _k^{{\mathrm{Re}} }}} = \frac{{\partial {\cal L}}}{{\partial \alpha _k^{{\mathrm{Im}} }}} = \frac{{\partial {\cal L}}}{{\partial {\lambda _1}}} = \frac{{\partial {\cal L}}}{{\partial {\lambda _2}}} = 0
\EE
and generates the following equation system
\begin{subequations}
	\begin{align}
	\label{eqS310a}
	{{\mathbf{A}}^{{\mathrm{GEP}}}}{\boldsymbol{\alpha }} &= {\lambda _1}{{\mathbf{C}}^{{\mathrm{GEP}}}}{\boldsymbol{\alpha }} + {\lambda _2} {{\mathbf{B}}^{{\mathrm{GEP}}}}{\boldsymbol{\alpha }}, \\
	\label{eqS310b}
	{{\boldsymbol{\alpha }}^{\mathrm{H}}}{{\mathbf{B}}^{{\mathrm{GEP}}}}{\boldsymbol{\alpha }} &= 1, \\
	\label{eqS310c}
	{{\boldsymbol{\alpha }}^{\mathrm{H}}}{{\mathbf{C}}^{{\mathrm{GEP}}}}{\boldsymbol{\alpha }} &= \gamma.
	\end{align}
\end{subequations}

When solving (\ref{eqS310a})--(\ref{eqS310c}), it is important to realize that (\ref{eqS310a}) is invariant with respect to scaling by a constant. The solution proceeds as follows:

\begin{itemize}
	\item Choose $\lambda_2$ and solve (\ref{eqS310a}).
	\item Normalize all solutions to satisfy (\ref{eqS310b}).
	\item Check the constraint (\ref{eqS310c}).
	\item Vary $\lambda_2$ and find solutions to (\ref{eqS310c}).
	\item From solutions satisfying (\ref{eqS310a})--(\ref{eqS310c}) select the one which minimizes the goal (\ref{eqS36a}).
\end{itemize}

Clearly, the constraints (\ref{eqS310b}) and (\ref{eqS310c}) can be mutually exclusive and the solution may not exist. In the special case of \mbox{$\gamma=0$}, which is the case of this paper, a solution always exists, as both (\ref{eqS310a}) and (\ref{eqS310c}) are invariant with respect to scaling by a constant. In this special case, the Lagrange multiplier $\lambda_2$ also attains clear meaning, for multiplying (\ref{eqS310a}) by $\boldsymbol{\alpha}^{\mathrm{H}}$ from the left and employing the constraints (\ref{eqS310b})--(\ref{eqS310c}) we obtain (for $\gamma=0$)
\BE
\label{eqS311}
{\lambda _2} = {{\boldsymbol{\alpha }}^{\mathrm{H}}}{{\mathbf{A}}^{{\mathrm{GEP}}}}{\boldsymbol{\alpha }},
\EE
i.e., in this case $\lambda_2$ is equal to the optimized function values. It is also important to realize that by a particular choice of matrices $\mathbf{S}$ and $\mathbf{T}$ we can always diagonalize at least two of the matrices ${{\mathbf{A}}^{{\mathrm{GEP}}}}$, ${{\mathbf{B}}^{{\mathrm{GEP}}}}$, ${{\mathbf{C}}^{{\mathrm{GEP}}}}$. The choice of matrices $\mathbf{S}$ and $\mathbf{T}$ is discussed in Section~\ref{results} and Section~\ref{disc}.

\section{Results}
\label{results}
The utility of the procedure presented in Section~\ref{optim} will now be presented through several examples which are of interest to designers of electrically small antennas. In each case we set \mbox{$\mathbf{C}=\mathbf{X}$} and \mbox{$\gamma=0$}, i.e., we force the resulting current to be resonant, a natural choice for antenna applications. The resonant assumption has two important implications. First, the optimization problem (\ref{eqS36a})--(\ref{eqS36c}) is not convex \cite{NocedalWright_NumericalOptimization}. Second, we are not forced to use posterior tuning by external lumped element, which, as will be shown, is not always the optimal choice.

\subsection{Maximization of the gain to quality factor ratio $G/Q$}
\label{GQopt}
The optimization of the $G/Q$ ratio is used as a proof of concept since its upper bound is known for electric current on arbitrarily shaped surfaces, see \cite{Gustafsson_OptimalAntennaCurrentsForQsuperdirectivityAndRP} and the references therein. Within the scheme presented in Section~\ref{optim}, the $G/Q$ ratio optimization induces \mbox{$\mathbf{A}=\omega\mathbf{W}$}, \mbox{$\mathbf{B}=4\pi\mathbf{U}$}. The optimal choice of matrices $\mathbf{S}$ and $\mathbf{T}$ will be discussed at a later point. At the present moment we have, quite arbitrarily, chosen \mbox{$\mathbf{S}=\mathbf{X}$} and \mbox{$\mathbf{T}=\omega\mathbf{W}$}, which leads to expansion modes defined via
\BE
\label{GQopt01}
{\mathbf{X}}{{\mathbf{I}}_n} = {\zeta _n}\omega{\mathbf{W}}{{\mathbf{I}}_n}.
\EE
This choice not only helps with numerical stability, as, together with the normalization (\ref{eqS34a}), it provides us with ${{\mathbf{A}}^{{\mathrm{GEP}}}}$ as the unit matrix and \mbox{${{\mathbf{C}}^{{\mathrm{GEP}}}}=\left[ {{\mathrm{diag}}\left( {{\zeta _n}} \right)} \right]$}, but it also gives us a clear indication of how to satisfy the constraint (\ref{eqS310c}) through recasting it to
\BE
\label{GQopt02}
	{{\boldsymbol{\alpha }}^{\mathrm{H}}}\left[ {{\mathrm{diag}}\left( {{\zeta _n}} \right)} \right]{\boldsymbol{\alpha }}= 0.
\EE
Recall that resonance, i.e., $\gamma = 0$ in (\ref{eqS310c}) is assumed. To satisfy the constraint (\ref{GQopt02}), the eigen-numbers $\zeta _n$ must\footnote{Equation~(\ref{GQopt02}) must be strictly adhered to. It can, however, happen that infinitesimally localized modes representing lumped circuit elements are needed for the optimal solution, see Section~\ref{disc}. This is, of course, only approximately possible with finite discretization.} be zero or contain both signs. Since $\zeta_n < 0$ implies 
\mbox{$\mathbf{I}^{\mathrm{H}}_n {\mathbf{X}}{{\mathbf{I}}_n}<0$} (excess electric energy) and \mbox{$\zeta_n > 0$} implies 
\mbox{$\mathbf{I}^{\mathrm{H}}_n {\mathbf{X}}{{\mathbf{I}}_n}>0$} (excess magnetic energy), this requirement means that the solution must be formed by a mixture of capacitive and inductive modes or, in exceptional cases, by modes in resonance, a natural requirement once we desire a resonant current.
The Lagrange multiplier $\lambda_2$ is equal to the $Q/G$ ratio. The solution of the system (\ref{eqS310a})--(\ref{eqS310c}) has been implemented in Matlab \cite{matlab} via generalized Schur decomposition \cite{Wilkinson_AlgebraicEigenvalueProblem} and a root-search method. The routine sought for the lowest value of $\lambda_2$ for which the system (\ref{eqS310a})--(\ref{eqS310c}) was satisfied.
\BF
	\begin{center}
		\includegraphics[width=8.9cm]{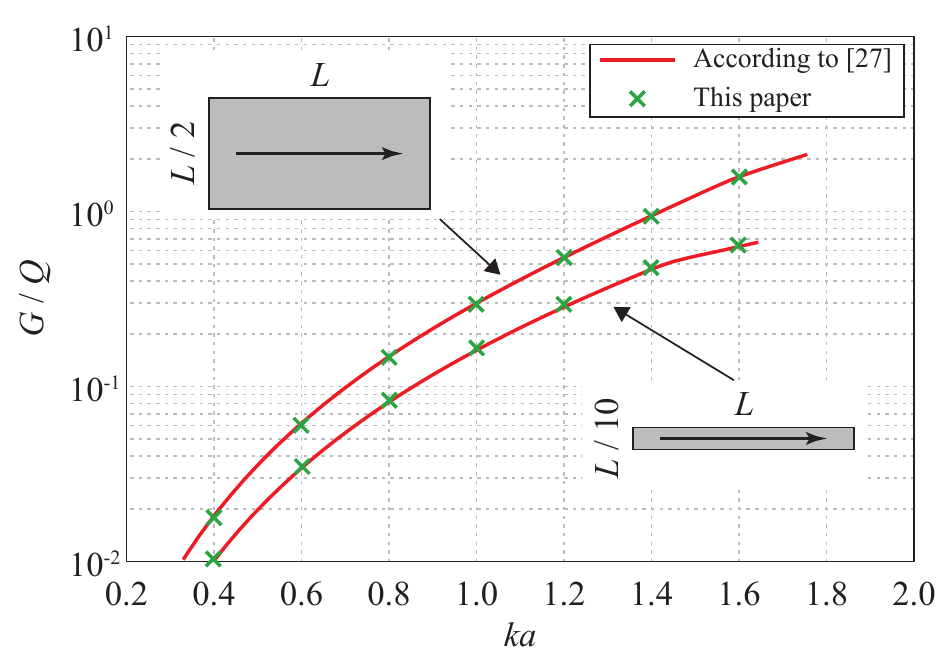}
		\caption{The upper bound of the $G/Q$ ratio for a rectangular patch as presented in \cite{GustafssonTayliEhrenborgEtAl_AntennaCurrentOptimizationUsingMatlabAndCVX} (full lines). The highest value of the $G/Q$ ratio found by the presented method (cross-marks). In both cases, the partial directivity in the direction normal to the patch is used with polarization along the longer edge of the patch (denoted by thick arrows). Parameter $a$ denotes the radius of the smallest circumscribing sphere.}
		\label{fig1}
	\end{center}
	\EF
A particular result of the $G/Q$ ratio maximization is presented in Fig.~\ref{fig1}. Apart from small discrepancies attributable to the discretization (approx. 670 triangles), the method presented in Section~\ref{optim} gives the same values as the upper bound found by the convex optimization \cite{GustafssonTayliEhrenborgEtAl_AntennaCurrentOptimizationUsingMatlabAndCVX} or upper bound found from the polarisability matrices \cite{GustafssonSohlKristensson_PhysicalLimitationsOfAntennasOfArbitraryShape_RoyalSoc}.

The current maximizing the $G/Q$ ratio corresponding to \mbox{$ka=0.4$} is depicted in Fig.~\ref{fig2} for the rectangular patch of proportions \mbox{$L \times L/2$}. For presentation purposes, the amplitude of the final current is further renormalized so as to radiate $1\;\mathrm{W}$. In this calculation, 41 modes of (\ref{GQopt01}) have been taken. The modal amplitudes of the final mixture are depicted in Fig.~\ref{fig3},
\begin{figure}[t]
\begin{center}
	\includegraphics[width=8.1cm]{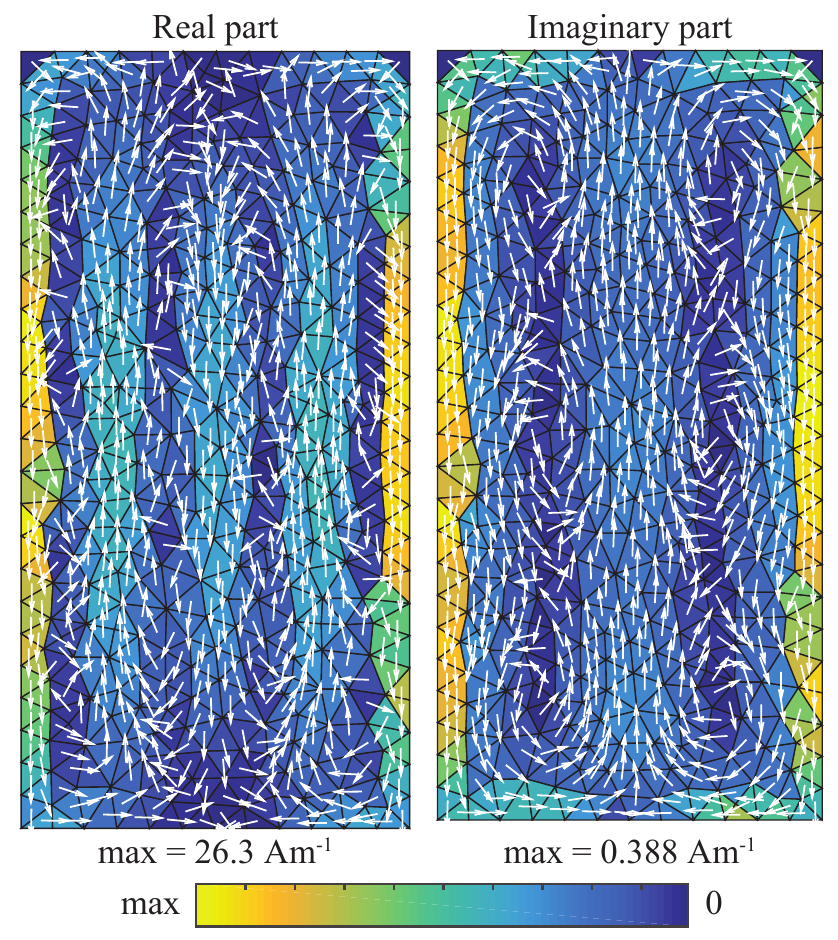}
	\caption{The current on a rectangular patch maximizing the $G/Q$ ratio at \mbox{$ka=0.4$}. The dimensions of the patch are \mbox{$1\;\mathrm{m} \times 0.5\;\mathrm{m}$} and the current is normalized so as to radiate $1\;\mathrm{W}$. A mix of 41 modes depicted in Fig.~\ref{fig3} was used.}
	\label{fig2}
\end{center}
\EF
\begin{figure}[t]
\begin{center}
\includegraphics[width=8.9cm]{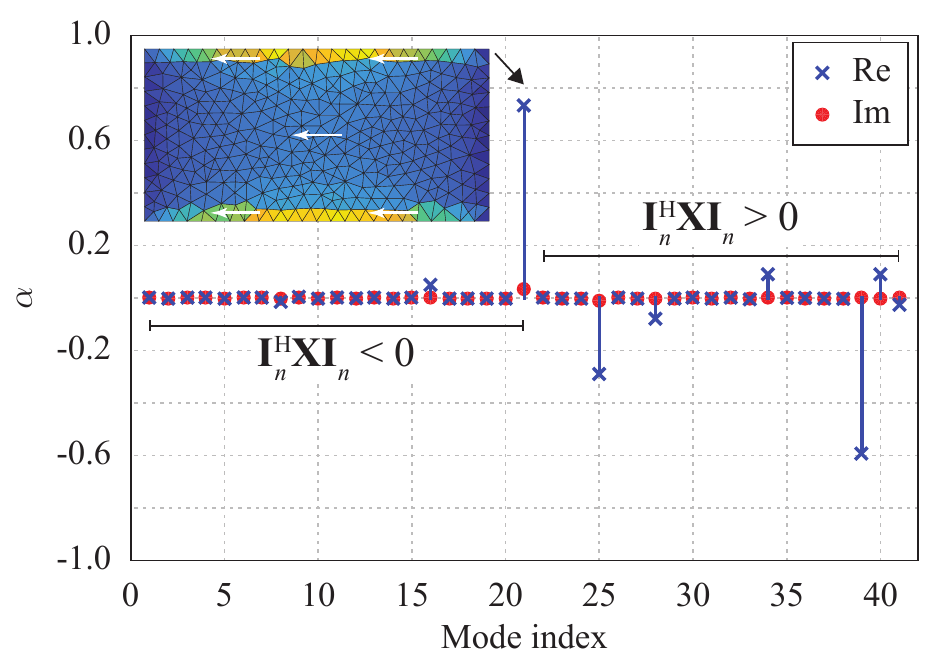}
\caption{A modal mixture of 41 modes corresponding to the optimal current depicted in Fig.~\ref{fig2}. The modes are ordered so that the negative (capacitive) values of $\mathbf{I}^{\mathrm{H}}_n {\mathbf{X}}{{\mathbf{I}}_n}$ grow in amplitude to the left, while positive (inductive) values of $\mathbf{I}^{\mathrm{H}}_n {\mathbf{X}}{{\mathbf{I}}_n}$ grow in amplitude to the right.}
\label{fig3}
\end{center}
\EF
in which the modes are ordered so that the negative (capacitive) eigenvalues $\zeta_n$ grow in amplitude to the left, while positive (inductive) eigenvalues grow in amplitude to the right. When selecting the modes for optimization, it is advantageous to chose those with high values of modal $G/Q$ ratio. Such modes are located at low magnitudes of $\zeta_n$. In this example, 21 capacitive modes, with the lowest magnitude of $\zeta_n$, and 20 inductive modes, with the lowest magnitude of $\zeta_n$, were taken. While the addition of more modes to the optimization process only has an insignificant influence on the optimal value of the $G/Q$ ratio, it does exhibit unexpected behaviour. The mixture of capacitive modes is stable, though the mixture of inductive modes changes dramatically as does the shape of the optimal current. This behaviour is depicted in Fig.~\ref{fig4} and Fig.~\ref{fig5}, where more inductive modes are used.
\begin{figure}[b]
\begin{center}
\includegraphics[width=8.9cm]{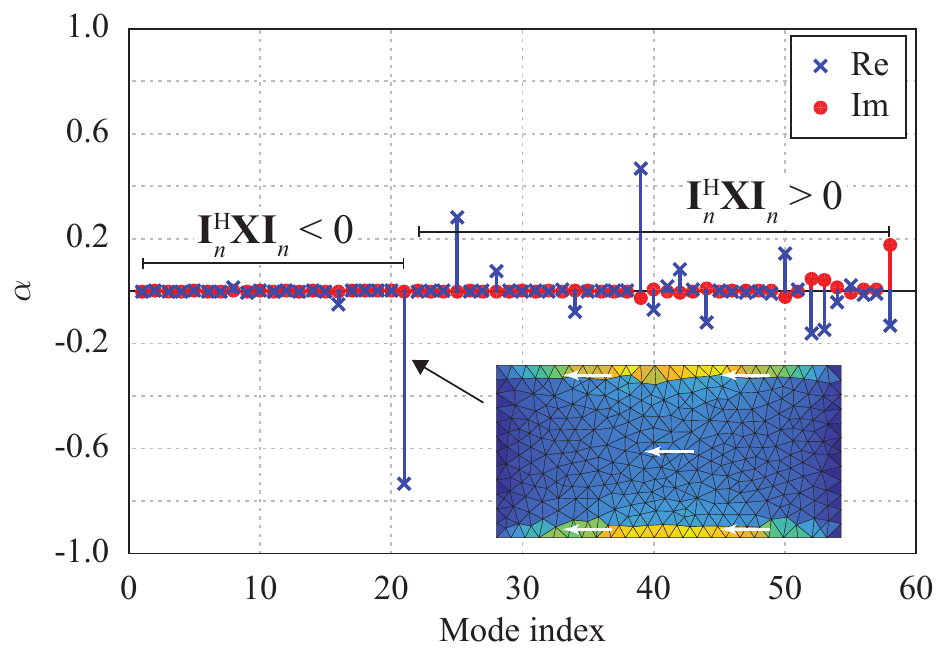}
\caption{A modal mixture of 58 modes corresponding to the optimal current depicted in Fig.~\ref{fig5}. The modes are ordered in the same manner as in Fig.~\ref{fig3}.}
\label{fig4}
\end{center}
\EF
These results suggest that convergence may not have been reached, a conclusion which is in sheer contradiction with Fig.~\ref{fig1}.
\BF
\begin{center}
\includegraphics[width=8.1cm]{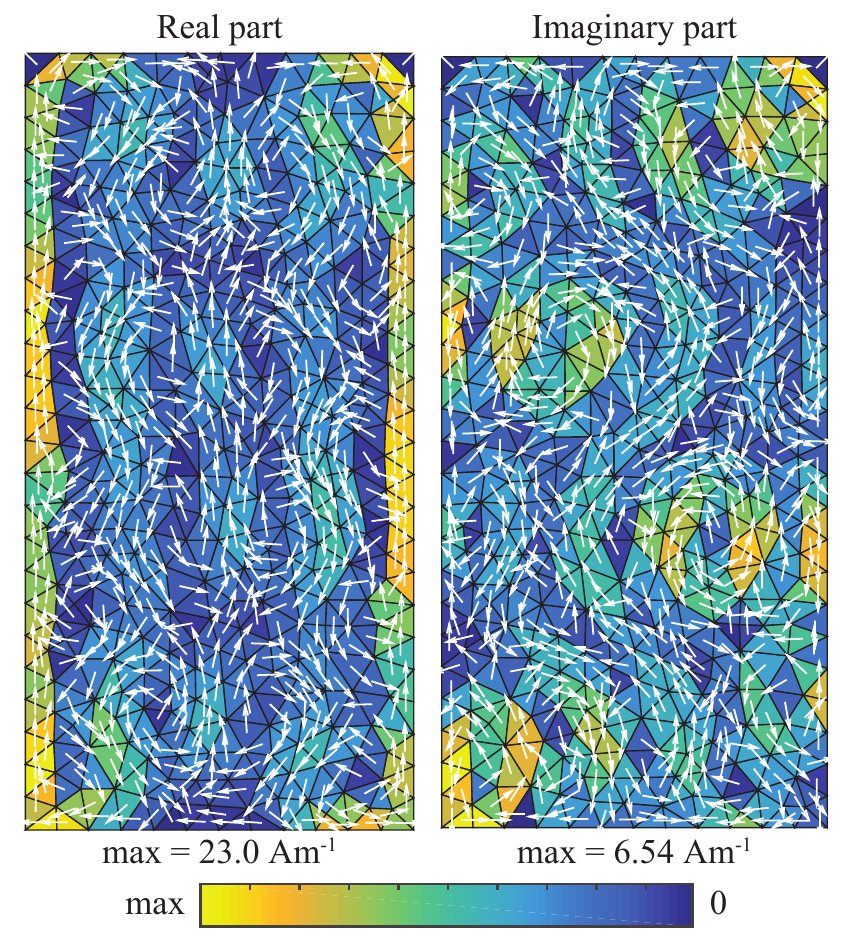}
\caption{The current on a rectangular patch maximizing the $G/Q$ ratio at \mbox{$ka=0.4$}. The dimensions of the patch are \mbox{$1\;\mathrm{m} \times 0.5\;\mathrm{m}$} and the current is normalized so as to radiate $1\;\mathrm{W}$. A mix of 58 modes depicted in Fig.~\ref{fig4} was used.}
\label{fig5}
\end{center}
\EF
The reason behind this discrepancy is given in \cite{GustafssonSohlKristensson_PhysicalLimitationsOfAntennasOfArbitraryShape_RoyalSoc,YaghjianStuart_LowerBoundOnTheQofElectricallySmallDipoleAntennas,GustafssonCismasuJonsson_PhysicalBoundsAndOptimalCurrentsOnAntennas_TAP} where it is shown that electric dipole moment guides the values of the $G/Q$ ratio at small electrical sizes and that adding loop-like currents with zero divergence does not affect the $G/Q$ ratio values while dramatically changing the current shape. The optimal current is not unique in this case, even though the upper bound of the $G/Q$ ratio (dictated solely by charge distribution) is \cite{GustafssonSohlKristensson_PhysicalLimitationsOfAntennasOfArbitraryShape_RoyalSoc,GustafssonCismasuJonsson_PhysicalBoundsAndOptimalCurrentsOnAntennas_TAP}.

The magnetization loop-like currents are exactly the modes added in Fig.~\ref{fig4}. However, in both presented cases, the charge distribution (given mostly by capacitive modes) is alike, with a shape according to Fig.~\ref{fig6} (both in real and imaginary part). The charge distribution evidently maximizes the electric dipole moment.
\BF
\begin{center}
\includegraphics[width=8.0cm]{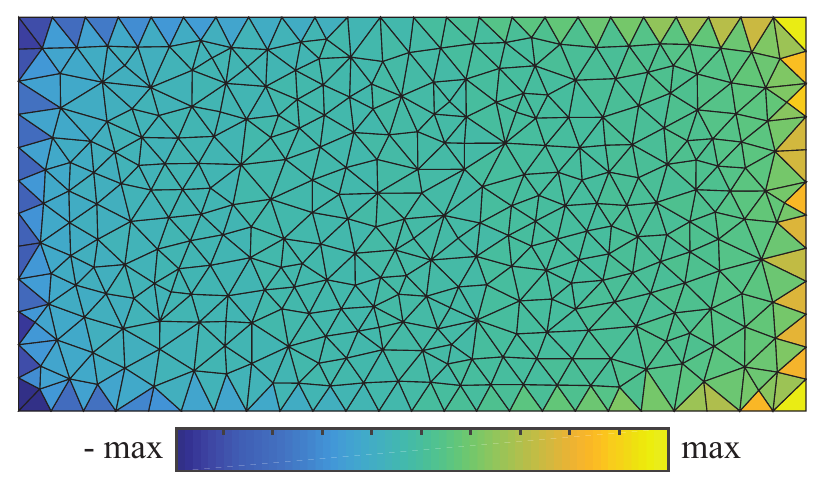}
\caption{Shape of the charge distribution corresponding to the optimal $G/Q$ ratio.}
\label{fig6}
\end{center}
\EF

The above discussion suggests that the maximization of the $G/Q$ ratio has too much freedom. Although the currents depicted in Fig.~\ref{fig2} and Fig.~\ref{fig5} are identical with respect to the $G/Q$ ratio, they certainly have a different quality factor $Q$ and different gain $G$, parameters that are of interest on their own. It is also noteworthy that the $G/Q$ ratio completely ignores losses, so adding a lossy constraint ${{\mathbf{I}}^{\mathrm{H}}}{\mathbf{\Sigma I}} < \kappa$ would be an interesting option.

\subsection{Minimization of quality factor $Q$}
\label{Qopt}
The minimization of quality factor $Q$ is a more challenging task which, in fact, has not yet been solved for arbitrarily shaped surfaces, although close approximations \cite{JonssonGustafsson_StoredEnergiesInElectricAndMagneticCurrentDensities_RoyA,Kim_LowerBoundsOnQForFinizeSizeAntennasOfArbitraryShape,ChalasSertelVolakis2016EarlyAccess,CapekJelinek_OptimalCompositionOfModalCurrentsQ} exist.

The formulation, via the method of Section~\ref{optim}, requires \mbox{$\mathbf{A}=\omega\mathbf{W}$}, \mbox{$\mathbf{B}={\mathbf{R}}/2$}. An interesting choice \mbox{$\mathbf{S}=\mathbf{X}$} and \mbox{$\mathbf{T}=\mathbf{R}$}, which is the same as in \cite{Kim_LowerBoundsOnQForFinizeSizeAntennasOfArbitraryShape,ChalasSertelVolakis2016EarlyAccess}, decomposes the solution into modes defined as
\BE
\label{Qopt01}
{\mathbf{X}}{{\mathbf{I}}_n} = {\zeta _n}{\mathbf{R}}{{\mathbf{I}}_n}
\EE
which are well-known characteristic modes \cite{MartaEva_TheTCMRevisited}. With these settings, the method searches for the lowest value of $\lambda_2$, which is directly equal to quality factor $Q$.

Two particular results for the minimum quality factor $Q$ on a rectangular patch and a disc are depicted in Fig.~\ref{fig7}. The optimal values of quality factor $Q$ are compared with the theoretical lower bound derived in \cite{McLean_AReExaminationOfTheFundamentalLimitsOnTheRadiationQofESA} which reads
\BE
\label{Qopt01a}
Q_{\mathrm{Chu}}^{\mathrm{TE} + \mathrm{TM}}=\frac{1}{2}\left(\frac{1}{\left(ka\right)^3} + \frac{2}{ka}   \right).
\EE
\BF
\begin{center}
\includegraphics[width=8.9cm]{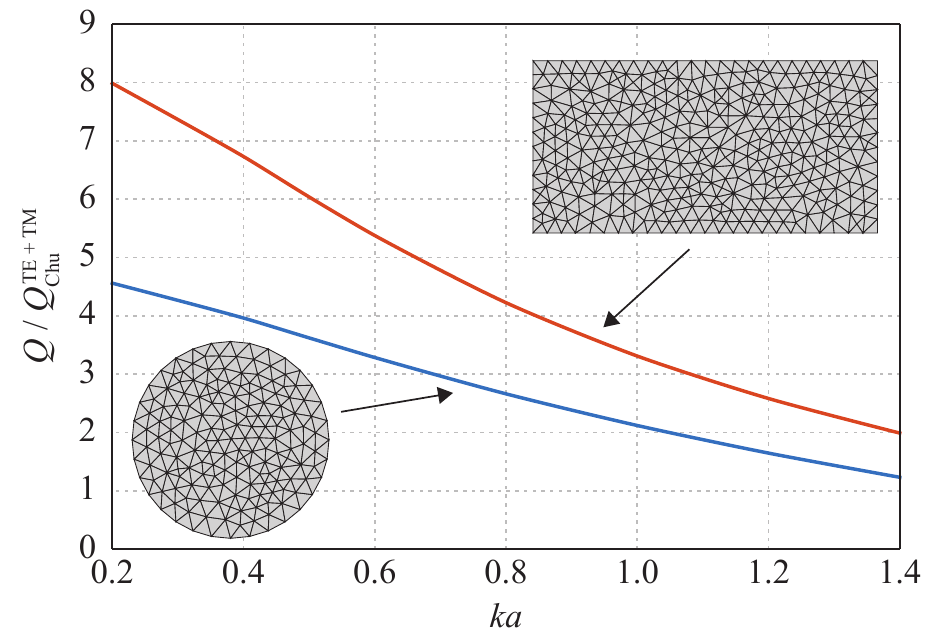}
\caption{The lower bound of quality factor $Q$ for a rectangular patch of proportions \mbox{$L \times L/2$} and for a disc of radius $a$. The radius $a$ also denotes the radius of the smallest circumscribing sphere. The values of quality factor $Q$ are normalized to the theoretical lower bound (\ref{Qopt01a}) derived in \cite{McLean_AReExaminationOfTheFundamentalLimitsOnTheRadiationQofESA}.}
\label{fig7}
\end{center}
\EF

Choosing the specific electrical size $ka=0.4$, Fig.~\ref{fig8} shows the current density exhibiting the minimum quality factor $Q$. In contrast to Section~\ref{GQopt}, Fig.~\ref{fig9} reveals that, visually, only two characteristic modes form the solution. Numerical tests show that with growing electrical size, more than two modes contribute, but their impact on the optimal value of quality factor $Q$ is minor. Clearly, the characteristic modes are a favourable basis for this optimization task as envisaged previously in \cite{Kim_LowerBoundsOnQForFinizeSizeAntennasOfArbitraryShape,ChalasSertelVolakis_ComputationOfTheLimitsForASAusingCM,ChalasSertelVolakis2016EarlyAccess}.
\BF
\begin{center}
\includegraphics[width=8.1cm]{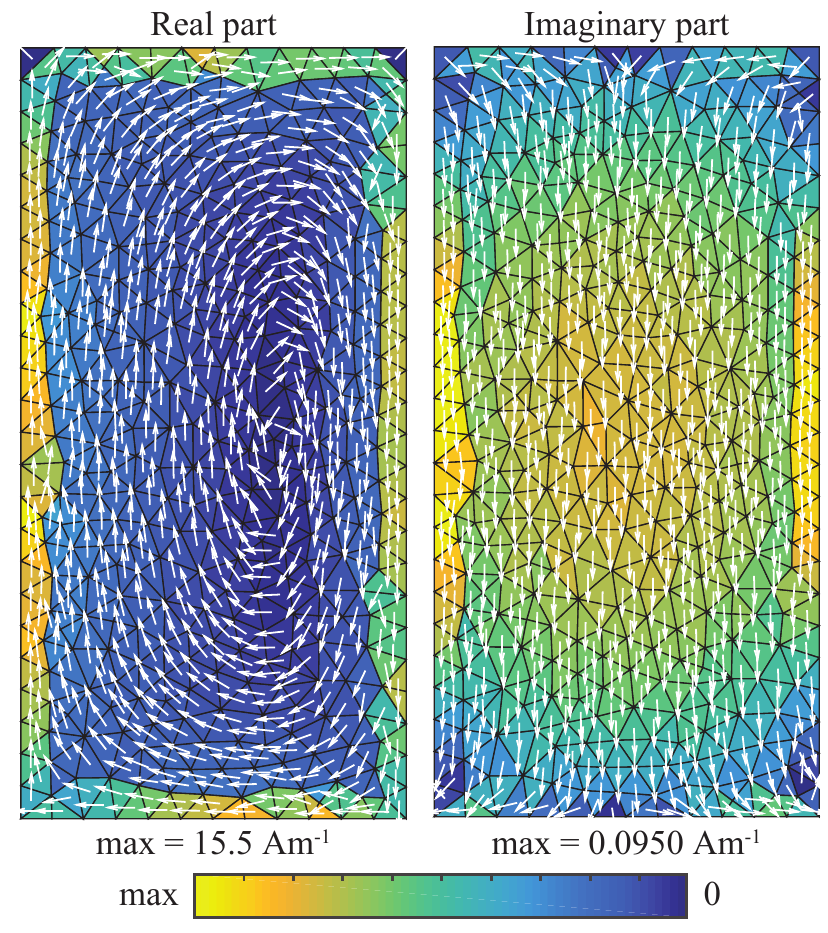}
\caption{The current on a rectangular patch minimizing quality factor $Q$ at \mbox{$ka=0.4$}. The dimensions of the patch are \mbox{$1\;\mathrm{m} \times 0.5\;\mathrm{m}$} and the current is normalized to radiate $1\;\mathrm{W}$. A mix of 11 modes depicted in Fig.~\ref{fig9} was used.}
\label{fig8}
\end{center}
\EF
\BF
\begin{center}
\includegraphics[width=8.9cm]{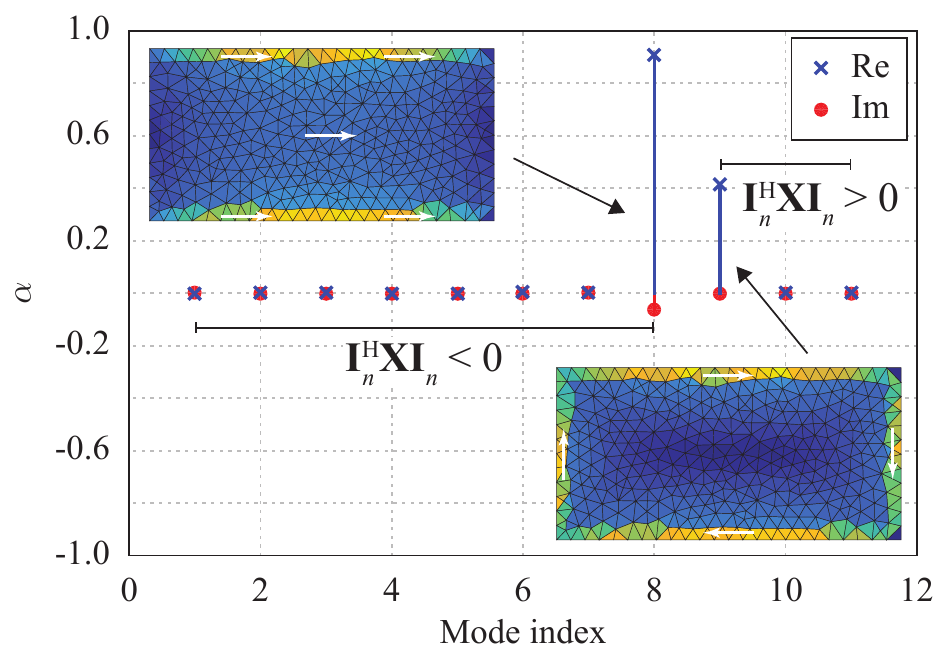}
\caption{A modal mixture of 11 modes corresponding to the optimal current depicted in Fig.~\ref{fig8}. The modes are ordered so that the negative (capacitive) values of $\mathbf{I}^{\mathrm{H}}_n {\mathbf{X}}{{\mathbf{I}}_n}$ grow in amplitude to the left, while positive (inductive) values of $\mathbf{I}^{\mathrm{H}}_n {\mathbf{X}}{{\mathbf{I}}_n}$ grow in amplitude to the right.}
\label{fig9}
\end{center}
\EF

The low number of characteristic modes necessary to achieve current minimizing quality factor $Q$ rises a question of whether the choice of matrices $\mathbf{S}$, $\mathbf{T}$ exists that will allow the optimal current to be composed of minimum number of modes. It can be shown that choosing
\mbox{$\mathbf{S}=\omega\mathbf{W}$}, \mbox{$\mathbf{T}=\mathbf{R}/2$}, i.e., to expand a solution into modes defined as
\BE
\label{Qopt02}
\omega\mathbf{W}{{\mathbf{I}}_n} =\frac{\zeta _n}{2}{\mathbf{R}}{{\mathbf{I}}_n},
\EE
is the desired choice as the modes (\ref{Qopt02}) can be made orthogonal, both with respect to matrix $\omega\mathbf{W}$ and $\mathbf{R}$. Assuming, thereby, that coefficients $\alpha_n$ have been set so as to make the current self-resonant (constraint (\ref{eqS31c}), (\ref{eqS36c})), the quality factor can be written as
\BE
\label{Qopt03}
Q = 2\omega \frac{{\sum\limits_n {{{\left| {{\alpha _n}} \right|}^2}{\mathbf{I}}_n^{\mathrm{H}}{\mathbf{W}}{{\mathbf{I}}_n}} }}{{\sum\limits_n {{{\left| {{\alpha _n}} \right|}^2}{\mathbf{I}}_n^{\mathrm{H}}{\mathbf{R}}{{\mathbf{I}}_n}} }} = \frac{{\sum\limits_n {{{\left| {{\alpha _n}} \right|}^2}{Q_n^{\mathrm{untuned}}}} }}{{\sum\limits_n {{{\left| {{\alpha _n}} \right|}^2}} }},
\EE
where the last equality assumes (\ref{eqS34a}) and where eigenvalues $\zeta _n$ given by (\ref{Qopt02}) have been renamed to $Q_n^{\mathrm{untuned}}$ as Eq. (\ref{Qopt02}) is giving them the meaning of untuned quality factors \cite{YaghjianBest_ImpedanceBandwidthAndQOfAntennas} of the stand-alone modes. Since both matrices $\omega\mathbf{W}$ and $\mathbf{R}$ are positively semi-definite (assuming infinite numerical precision and neglecting the rare possibility of the negative values of \mbox{${{\mathbf{I}}_{n}^{\mathrm{H}}}\omega\mathbf{W}{{\mathbf{I}}_n}$} \cite{GustafssonCismasuJonsson_PhysicalBoundsAndOptimalCurrentsOnAntennas_TAP}), the eigenvalues \mbox{$\zeta_n=Q_n^{\mathrm{untuned}}>0$}. The summations in (\ref{Qopt03}) can, thus, only grow with an increasing number of terms. Ordering the summation so that \mbox{$\zeta_n=Q_n^{\mathrm{untuned}}$} is a growing sequence, the minimum quality factor $Q$ is closely approached either by the first term, if it is self-resonant, or by a combination of capacitive and inductive modes with the lowest \mbox{$\zeta_n=Q_n^{\mathrm{untuned}}$}. The formula (\ref{Qopt03}) also shows that \mbox{$Q \ge Q_1^{\mathrm{untuned}}$}.

Using the above approximation by two modes, the optimization task leading to the resonant current minimizing quality factor $Q$ is analytically solvable, once the modes of (\ref{Qopt02}) are known. The solution is to take a mode (\ref{Qopt02}) with the lowest eigenvalue and mix it with a second mode which has the opposite sign of ${{\mathbf{I}}_{n}^{\mathrm{H}}}\mathbf{X}{{\mathbf{I}}_n}$ with respect to the first mode and has the lowest possible eigenvalue, a conclusion also reached in \cite{CapekJelinek_OptimalCompositionOfModalCurrentsQ} by a different methodology. Addition of more modes commonly presents only slight improvement.

Although straightforward, care should be taken when implementing the above procedure. The issue comes from the notoriously ill-conditioned matrix $\mathbf{R}$ \cite{CapekHazdraEichler_ComplexPowerRatioFunctionalForRadiatingStructures}, which results in only a few modes of (\ref{Qopt01}), (\ref{Qopt02}) being numerically stable on electrically small structures. Common algorithms, such as the generalized Schur decomposition or the implicitly restarted Arnoldi method as implemented in the Matlab \cite{matlab}, are often unable to generate modes satisfying (\ref{eqS34a}) which ruins the theoretical reasoning below (\ref{Qopt03}). It is highly advisable to utilize some of the multi-precision packages, such as \cite{advanpix}, for the eigenvalue decomposition if attempts are made to obtain the true global optimum in the basis (\ref{Qopt02}). Numerical tests suggest that quad precision is satisfactory for the mesh densities presented.

\subsection{Maximization of gain $G$}
\label{Gopt}
Although the current in the prescribed region can always be formed so as to have an infinitely sharp radiation pattern and, therefore, infinite directivity \cite{Bloch_superdirectivity_1953}, the consequences are dire. First, the quality factor $Q$ of such superdirective current is extremely high, making the ratio $D/Q$ much below the optimum shown in Section~\ref{GQopt}. Second, the superdirective current gives rise to high losses, making gain $G$ finite and not optimal \cite{max_gain_2012_Arbabi_TAP}. 

In this section, the current reaching the optimal value of gain $G$ is presented. Within the scheme shown in Section~\ref{optim}, the optimization task is characterized by \mbox{$\mathbf{A}=\left(\mathbf{R}+\mathbf{\Sigma}\right)/2$}, \mbox{$\mathbf{B}=4\pi{\mathbf{U}}$}. Using furthermore the results of 
Section~\ref{Qopt}, an interesting choice \mbox{$\mathbf{S}=4\pi\mathbf{U}$} and \mbox{$\mathbf{T}=\left(\mathbf{R}+\mathbf{\Sigma}\right)/2$} decomposes the solution into the modes of
\BE
\label{Gopt01X}
{4\pi\mathbf{U}}{{\mathbf{I}}_n} = \frac{\zeta_n}{2}{\left(\mathbf{R}+\mathbf{\Sigma}\right)}{{\mathbf{I}}_n},
\EE
which eigenvectors diagonalize matrices $\mathbf{A}$ and $\mathbf{B}$ and which eigenvalues are equal to gain $G_n$ of stand-alone modes. 

The above choice of the basis has two important consequences. First, the fact that matrix $\mathbf{U}$ is composed of a single vector (see Section~\ref{appSigma}) leads to zero eigenvalues $\zeta_n$ with exception of one. Second, the product \mbox{$\mathbf{I}^{\mathrm{H}}_n {\mathbf{X}}{{\mathbf{I}}_n}$} has the same sign for all modes. Constructing, thus, an equation analogous to (\ref{Qopt03}) leaves us with a single non-zero term. The single non-zero eigenvalue is the value of the maximum achievable gain. Unfortunately, the structure cannot be brought to resonance since all modes are, for small electrical sizes, capacitive. This is, however, only true for finite discretization. If an infinite discretization were available, a mode (with zero eigenvalue) would exist simulating an ideal reactive lumped element leaving the value of gain untouched, but bringing the system to the resonance. The maximization of gain is thereby a trivial task requiring us only to find a non-zero eigenvalue of (\ref{Gopt01X}). Notice that this conclusion is identical to that reached for a multi-port antenna system \cite{UzsokySolymar_TheoryOfSuperDirectiveLinearArrays,Harrington_AntennaExcitationForMaximumGain}.

The optimal values of gain for current distributed on a spherical surface, a disc and rectangular patch are depicted in Fig.~\ref{fig10}. The results are normalized with respect to the so called ``normal gain'' \cite{Harrington_EffectsOfAntennaSizeOnGainBWandEfficiency} which reads
\BE
\label{Gopt02}
G_{\mathrm{normal}}=\left(ka\right)^2+2ka,
\EE
where $a$ is the radius of the smallest circumscribing sphere. The optimal gain was sought in the direction normal to the surface (irrelevant for a sphere) with polarization along the longer edge of the rectangular patch (irrelevant for a sphere and a disc). In order to make the problem scalable, the ratio \mbox{$\sigma /  \left(\omega \epsilon\right)$} has been fixed, rather than the value of conductivity $\sigma$. The current layer thickness is assumed to be much bigger than the penetration depth, see Section~\ref{appSigma}.

\BF
\begin{center}
	\includegraphics[width=8.9cm]{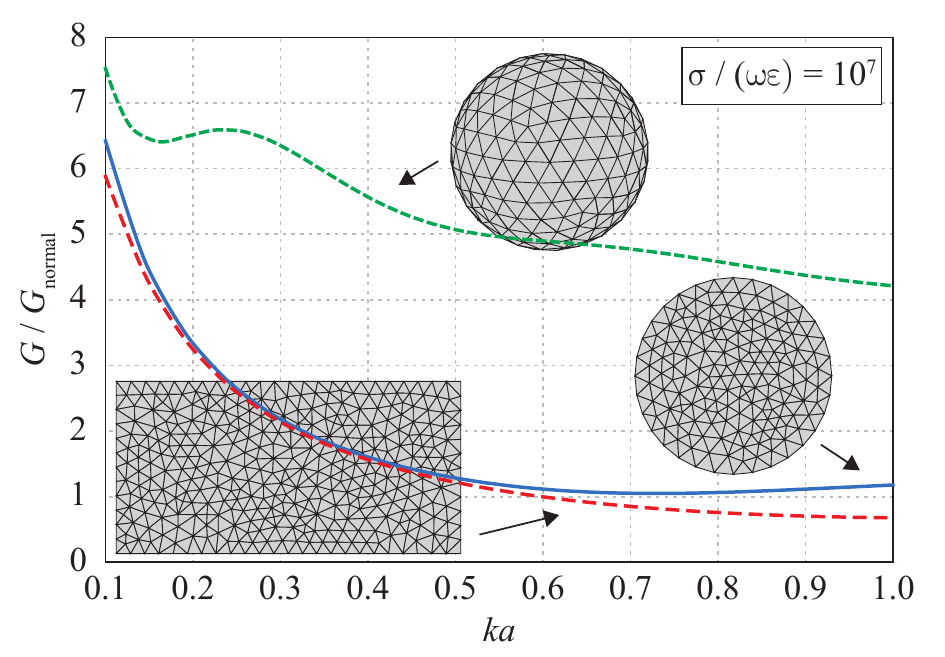}
	\caption{The upper bound of gain $G$ for a rectangular patch of proportions $L \times L/2$, for a disc of radius $a$, and for a spherical shell of radius $a$. Radius $a$ also denotes the radius of the smallest circumscribing sphere. The values of gain $G$ are normalized to the so called ``normal gain'' \cite{Harrington_TimeHarmonicElmagField,Chu_PhysicalLimitationsOfOmniDirectAntennas} defined by (\ref{Gopt02}). In order to make the problem scalable, the ratio $\sigma /  \left(\omega \epsilon\right)$ has been fixed.}
	\label{fig10}
\end{center}
\EF
The current leading to the optimal gain at $ka=0.4$ is depicted in Fig.~\ref{fig11} and the corresponding radiation pattern is shown in Fig.~\ref{fig12}. The radiation pattern bears no sign of superdirectivity and resembles that of an electric dipole oriented along the longer edge of the rectangular patch. 
\BF
\begin{center}
	\includegraphics[width=8.1cm]{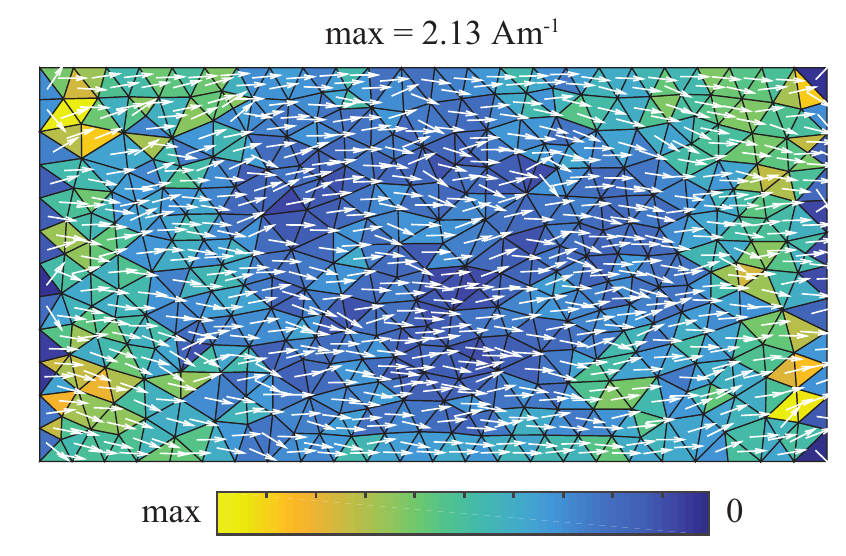}
	\caption{The current on a rectangular patch maximizing gain $G$ at $ka=0.4$ in the direction normal to the patch and with polarization along the longer edge of the rectangle. The dimensions of the patch are $1\;\mathrm{m} \times 0.5\;\mathrm{m}$ and the current is normalized so as to radiate $1\;\mathrm{W}$. The losses are assumed to be the same as in  Fig.~\ref{fig10}.}
	\label{fig11}
\end{center}
\EF

\BF
\begin{center}
	\includegraphics[width=7.1cm]{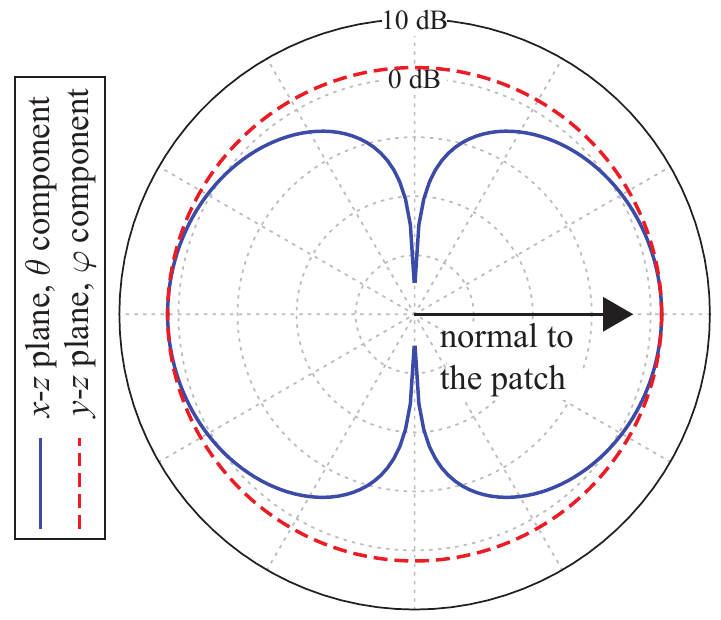}
	\caption{The partial directivity corresponding to the current density depicted in Fig.~\ref{fig11}. The coordinate origin coincides with the centre of the rectangle, the $z$-axis is normal to the rectangle and the $x$-axis is parallel to its longer edge.}
	\label{fig12}
\end{center}
\EF

\subsection{Maximization of radiation efficiency $\eta$}
\label{etaopt}

A search for the upper bound of radiation efficiency $\eta$ has the poorest history from the antenna metrics discussed so far. A rigorous upper bound is known for a current density filling a spherical volume \cite{Fujita_max_gain_APS_2013,2013_Karlsson_PIER,Fujita_max_gain_IEICE_TE_2015}, while only its estimate \cite{2016_Shahpari_Arxiv} exists for surfaces of arbitrary shape. This section therefore aims at giving a prescription for a current distribution maximizing radiation efficiency $\eta$ on an arbitrary surface.

Following the previous sections, we set \mbox{$\mathbf{A}=\left(\mathbf{R}+\mathbf{\Sigma}\right)$}, \mbox{$\mathbf{B}={\mathbf{R}}$}, and without hesitation we take \mbox{$\mathbf{S}=\mathbf{R}$} and \mbox{$\mathbf{T}=\left(\mathbf{R}+\mathbf{\Sigma}\right)$}, i.e., decompose the solution into the modes of
\BE
\label{etaopt01}
{\mathbf{R}}{{\mathbf{I}}_n} = {\zeta_n}{\left(\mathbf{R}+\mathbf{\Sigma}\right)}{{\mathbf{I}}_n}.
\EE
Note that the eigenvalues $\zeta_n$ of (\ref{etaopt01}) are equal to the modal radiation efficiencies.

Ensuring (\ref{eqS34a}), the total radiation efficiency is given by equation analogous to (\ref{Qopt03}), but unlike the minimization of quality factor $Q$ and similarly to the maximization of gain $G$, all the eigensolutions have the same sign of the product \mbox{$\mathbf{I}^{\mathrm{H}}_n {\mathbf{X}}{{\mathbf{I}}_n}$}, being all capacitive for small values of $ka$. This observation tells us that the optimal resonant current is given by the mode of (\ref{etaopt01}) with the highest radiation efficiency which is then tuned to resonance by a lumped element. Taking the best possibility of tuning by lossless lumped reactance, the upper bound of radiation efficiency $\eta$ for a current distributed on a spherical surface, on a disc and a rectangular patch are depicted by continuous lines in Fig.~\ref{fig13}, where the normalization of the losses is the same as in Section~\ref{Gopt}.

The results are also compared to the upper estimate \cite{2016_Shahpari_Arxiv}
\BE
\label{etaopt02}
\eta_{\mathrm{max}} = \left( 1+ 6 \pi \frac{\mathrm{Re} \left\{ {{Z_\mathrm{s}}} \right\}}{Z_0 k^2 S} \right)^{-1}
\EE
with $Z_0$ being the freespace impedance, $S$ being the total surface of the analysed object and $Z_\mathrm{s}$ being the surface impedance of a lossy conductor defined by (\ref{AppD01}) or (\ref{AppD03}). The results of (\ref{etaopt02}) are depicted by circle marks. Coherently with the exposition given in \cite{2016_Shahpari_Arxiv}, the prescription (\ref{etaopt02}) always lies above the true upper bound given by the method of this paper. For planar shapes presented in Fig.~\ref{fig13} the estimate (\ref{etaopt02}) must however be denoted as a close one.

\BF
\begin{center}
	\includegraphics[width=8.1cm]{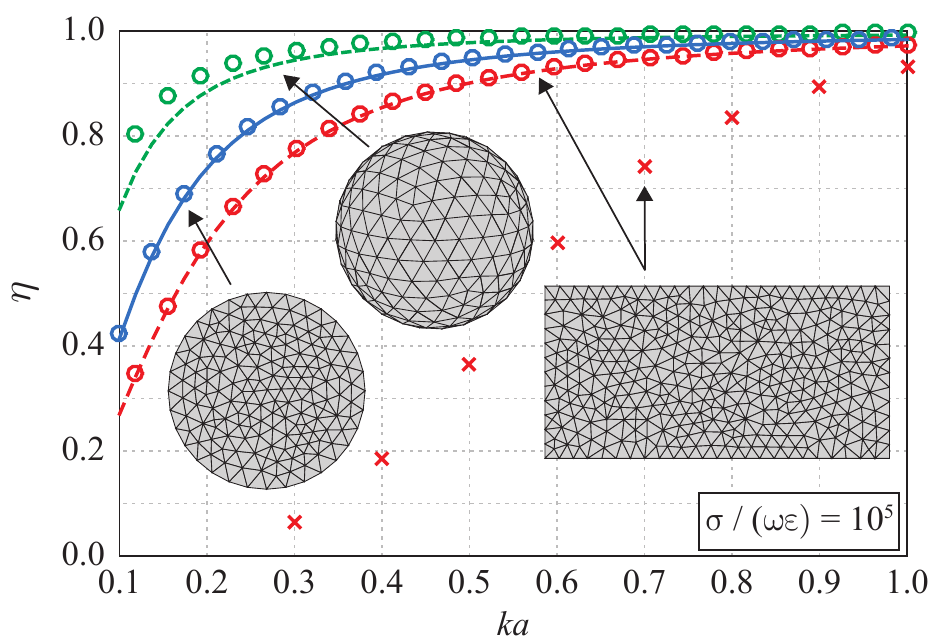}
	\caption{ The upper bound (continuous lines and cross marks) of radiation efficiency $\eta$ for a rectangular patch of proportions $L \times L/2$, for a disc of radius $a$ and for a sphere of radius $a$. Radius $a$ also denotes the radius of the smallest circumscribing sphere. The results are also compared to the estimate (\ref{etaopt02}) presented by circular marks.  In order to make the problem scalable, the ratio $\sigma /  \left(\omega \epsilon\right)$ has been fixed.}
	\label{fig13}
\end{center}
\EF

The lumped lossless tuning used above is however not realistic and an antenna designer could ask what is the achievable radiation efficiency $\eta$ when, as an example, only 20 characteristic modes with lowest magnitude of the eigenvalue (filtering our those exhibiting negative radiated power or complex eigenvalues) is available. The method presented in this paper is able to give the answer which is depicted by the cross marks in Fig.~\ref{fig13}. The solution is mostly generated by a mixture of capacitive and inductive mode with the lowest eigenvalue. The drop with respect to the true upper bound is considerable.

\section{Excitation of the optimal current}
\label{excit}
The optimal current obtained by the method presented in this paper sets an absolute bound to a given antenna metric for a given surface. It is important to stress that this current has to be seen as an impressed source in vacuum with no real support. If such a current is to be supported by a conducting surface, then a delta gap excitation would be needed at every discretization edge, as suggested by the discretized electric field integral equation (assuming no losses) \cite{Harrington_FieldComputationByMoM}
\BE
\label{eq51}
{\mathbf{V}}={\mathbf{Z}}{{\mathbf{I}}^{{\mathrm{opt}}}},
\EE
where ${{\mathbf{I}}^{{\mathrm{opt}}}}$ is the vector of edge current densities representing the optimal current in the RWG basis and $\mathbf{V}$ is the vector of excitation coefficients \cite{Harrington_FieldComputationByMoM}, which would be full of non-zero entries in this case. By reducing the number of feeding points, optimality always has to be sacrificed, notwithstanding the fact that using stand-alone feeders inside conducting regions will lead to short-circuits via their surroundings.

Based on the above discussion, the interesting question arises of how to cover parts of the optimized region by a conductor so that a good approximation of the optimal current would be excited by a user-defined number of feeders. In the current state of understanding, two options exist for solving this task, both of which are heuristic. The first option uses a defined number of feeding points and a heuristic algorithm, combined with a method, such as pixelling \cite{RahmatSamii_Kovitz_Rajagopalan-NatureInspiredOptimizationTechniques}, to optimize the positions and complex amplitudes of the feeders, as well as the structure of the supporting conductor. Unfortunately, it typically leads to a current notably different in shape to the optimal one \cite{ErentokSigmund2011,HassanWadbroBerggren_TopologyOptimizationOfMetallicAntennas,CismasuGustafsson_FBWbySimpleFreuqSimulation}, although exhibiting values close to the optimum for the metric at hand. The second option starts with finding the optimal current in the basis generated by characteristic modes (\ref{Qopt01}) with the hope they can selectively be excited \cite{MartensManteuffel_SystematicDesignMethodOfMobileMultipleAntennaSystemUsingCM}. At this point the optimization scheme presented in Section~\ref{optim} becomes extremely useful, since the characteristic modes are not the optimal basis for most of the optimization tasks. The presented optimization scheme, however, always returns the optimal modal composition regardless of the basis.

The search for an approximation of the optimal currents presented in the previous sections, but having realistic feeding, is out of the scope of this paper. In one particular example we will, nevertheless, show a trade-of between the number of feeders and optimality. For the example we choose the current density shown in Fig.~\ref{fig8} which minimizes quality factor $Q$ at $ka=0.4$ on a rectangular patch region. The optimal composition of characteristic modes ${{\mathbf{\alpha}}^{{\mathrm{opt}}}}$ corresponding to this current is depicted in Fig.~\ref{fig9}. 

The shape of the current depicted in Fig.~\ref{fig8} suggests that if the internal parts of the region are cut out, leaving only a thin strip coinciding with the external periphery, the optimal current should mostly be unharmed. This reasoning can be verified by a direct calculation of the optimal current on such a loop by the method described in Section~\ref{Qopt}. The result is depicted in Fig.~\ref{fig15} and Fig.~\ref{fig16}. The resulting quality factor equals $78.9$, while the current of Fig.~\ref{fig8} gives $Q=69.5$. The rise of quality factor $Q$ induced by the removal of the central region is to be expected, since the new structure has lower polarizability than a complete rectangle \cite{GustafssonSohlKristensson_PhysicalLimitationsOfAntennasOfArbitraryShape_RoyalSoc}. At this modest sub-optimality, we have, however, obtained a structure with many possible feeding edges.
\BF
\begin{center}
	\includegraphics[width=8.1cm]{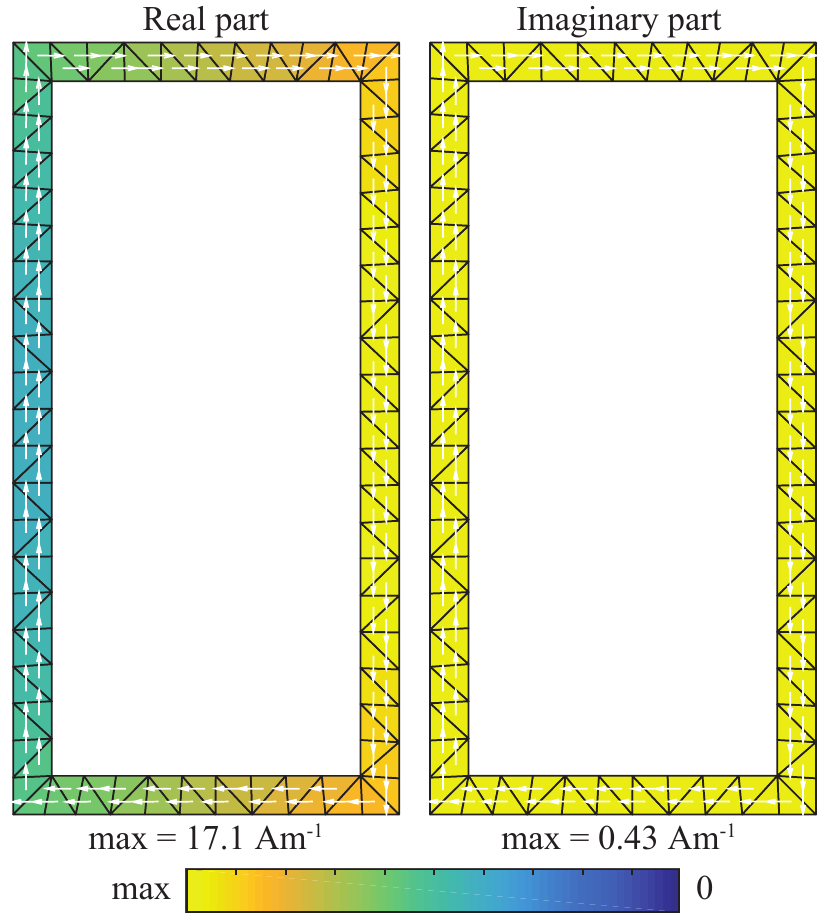}
	\caption{The current on a loop region minimizing quality factor $Q$ at \mbox{$ka=0.4$}. The outer dimensions of the loop are \mbox{$1\;\mathrm{m} \times 0.5\;\mathrm{m}$} and the current is normalized to radiate $1\;\mathrm{W}$. The width of the strip is equal to \mbox{$0.05\;\mathrm{m}$.}}
	\label{fig15}
\end{center}
\EF

\BF
\begin{center}
	\includegraphics[width=8.1cm]{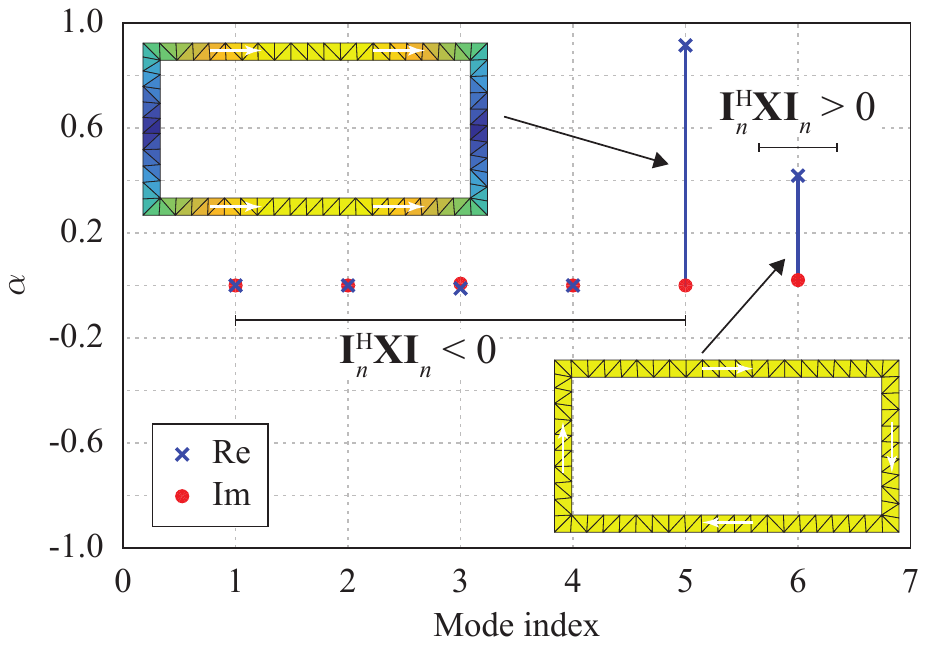}
	\caption{A modal mixture of 6 characteristic modes corresponding to the optimal current depicted in Fig.~\ref{fig15}. The modes are ordered so that the negative (capacitive) values of $\mathbf{I}^{\mathrm{H}}_n {\mathbf{X}}{{\mathbf{I}}_n}$ grow in amplitude to the left, while positive (inductive) values of $\mathbf{I}^{\mathrm{H}}_n {\mathbf{X}}{{\mathbf{I}}_n}$ grow in amplitude to the right.}
	\label{fig16}
\end{center}
\EF
Keeping the structure fixed and covering it with PEC, the question of how a given number of feeders should be distributed along the structure, and with what complex amplitudes, in order to achieve minimum quality factor $Q$ can be asked. We have approached this task through a genetic algorithm \cite{Deb_MultiOOusingEA}, the result of which is depicted in Fig.~\ref{fig17}.
\BF
\begin{center}
	\includegraphics[width=8.1cm]{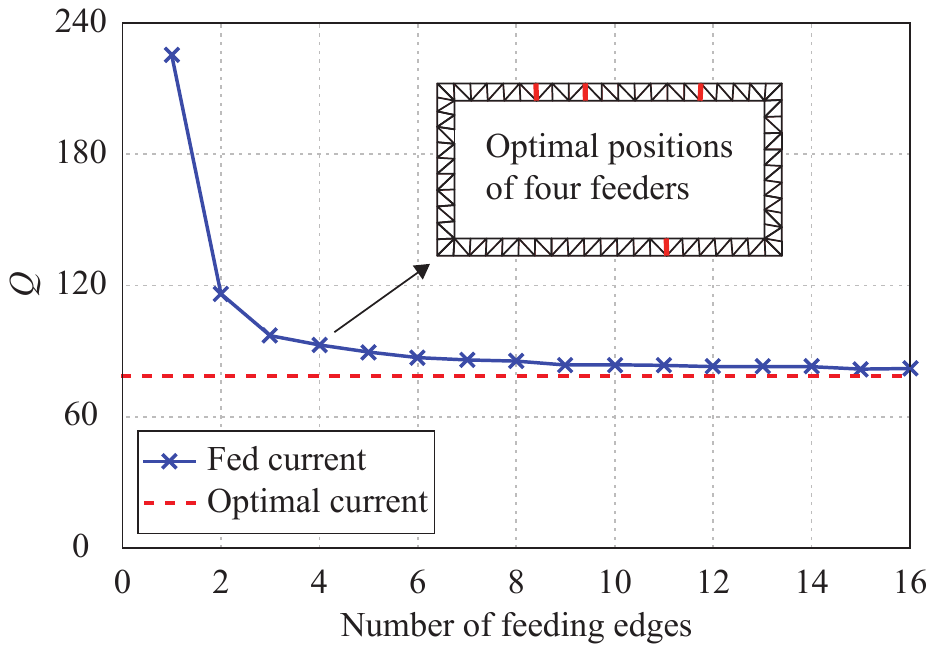}
	\caption{Comparison of quality factor $Q$ for the optimal current depicted in Fig.~\ref{fig15} and the current fed by a given number of feeding edges in the same domain covered with PEC. The current fed by a single feeder is not in resonance though other cases are. The full line connecting the markers is solely used for presentation purposes. The values of quality factor $Q$ between these points have no meaning. The inset shows the optimal positions of four feeding edges.}
	\label{fig17}
\end{center}
\EF

With respect to this result it is also interesting to know the modal composition of the fed current with respect to the six modes used for the optimal current, see Fig.~\ref{fig16}. Using the orthogonal properties of characteristic modes, the modal coefficients $\mathbf{\alpha}$ for a given feeding vector $\mathbf{V}$ can easily be calculated as \cite{HarringtonMautz_TheoryOfCharacteristicModesForConductingBodies}
\BE
\label{eq53}
\alpha_n =   \frac{{{\mathbf{I}}_n^{\mathrm{H}}} {\mathbf{V}}}{{1 + {\mathrm{j}}{\zeta_n}}}.
\EE
Ordering the modes in the same way as in Fig.~\ref{fig16}, the amplitudes of coefficients $\mathbf{\alpha}$ are depicted in Fig.~\ref{fig18} for all feeding options used in Fig.~\ref{fig17}.
\BF
\begin{center}
	\includegraphics[width=8.1cm]{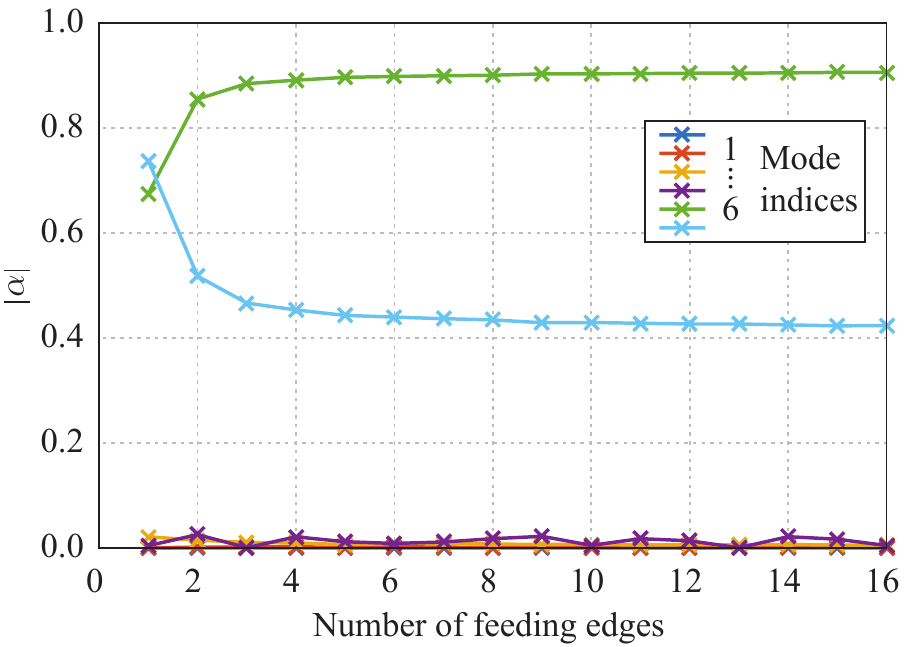}
	\caption{Excitation coefficients of characteristic modes corresponding to various feeding scenarios of Fig.~\ref{fig17}. The ordering of modes is the same as in Fig.~\ref{fig16}. The meaning of full lines is the same as in Fig.~\ref{fig17}.}
	\label{fig18}
\end{center}
\EF

\section{Discussion}
\label{disc}
This section considers some important properties of the proposed optimization scheme.

\subsection{Optimal choice of $\mathbf{S}$ and $\mathbf{T}$}

The results presented in Section~\ref{results} clearly show that the optimal way of solving (\ref{eqS31a})--(\ref{eqS31c}) is to represent the solution in the basis generated by
\BE
\label{eq41a}
{\mathbf{A}}{{\mathbf{I}}_n} = {\zeta _n}{\mathbf{B}}{{\mathbf{I}}_n},
\EE
or
\BE
\label{eq41b}
{\mathbf{B}}{{\mathbf{I}}_n} = {\zeta _n}{\mathbf{A}}{{\mathbf{I}}_n}.
\EE

Such a choice allow us to find the optimal current in terms of few modes of (\ref{eq41a}) or (\ref{eq41b}) so chosen as to have \mbox{minimum / maximum} possible eigenvalues $\zeta_n$, while allowing (\ref{eqS31c}) to be satisfied.

The optimal $G/Q$ ratio should be sought in the basis
\BE
\label{eq42}
{4\pi\mathbf{U}}{{\mathbf{I}}_n} = {\zeta _n}{\omega\mathbf{W}}{{\mathbf{I}}_n}
\EE
and tuned to resonance by a distributed current ${{\mathbf{I}}_n}$. The optimal quality factor $Q$ should be sought in the basis
\BE
\label{eq43}
{\omega\mathbf{W}}{{\mathbf{I}}_n} = \frac{\zeta_n}{2}{\mathbf{R}}{{\mathbf{I}}_n}
\EE
and tuned to resonance by a distributed current ${{\mathbf{I}}_n}$. The optimal gain $G$ should be sought in the basis
\BE
\label{eq44}
{4\pi\mathbf{U}}{{\mathbf{I}}_n} = \frac{\zeta_n}{2}{\left(\mathbf{R}+\mathbf{\Sigma}\right)}{{\mathbf{I}}_n}
\EE
and tuned to resonance by a lumped reactive element and, lastly, the optimal radiation efficiency should be sought in the basis
\BE
\label{eq45}
{\mathbf{R}}{{\mathbf{I}}_n} = {\zeta_n}{\left(\mathbf{R}+\mathbf{\Sigma}\right)}{{\mathbf{I}}_n}
\EE
and tuned to resonance by a lumped reactive element. 

\subsection{Other choices of $\mathbf{S}$ and $\mathbf{T}$ }
If other, less optimal bases, are used, the method presented in this paper finds the best solution within the number of expansion modes used. Enlarging the number of modes allows us to approach the optimal solution as closely as desired. The number of necessary modes can, however, be large, particularly when a metric, whose optimum is reached by a lumped circuit tuning, is being optimized, see Section~\ref{etaopt} for an example.

There could, nevertheless, be strong motivations for using non-optimal bases and sacrificing the optimality. One reason could be the inaccessibility of lumped tuning elements, another, an attempt for realistic excitation of the optimal current.

\subsection{Global optimum}
With $\gamma = 0$ in (\ref{eqS310c}), which is the case of all optimizations performed in this paper, the optimization process can always be formed so that the global optimum corresponds to the lowest value of the Lagrange multiplier $\lambda_2$ for which the system (\ref{eqS310a})--(\ref{eqS310c}) is satisfied, see Section~\ref{results} for examples. In principle, the method thus finds the global optimum. In reality one must be careful, as the slope of the constraint (\ref{eqS310c}) as a function of $\lambda_2$ can be steep.

\subsection{Excitation of the optimal current}
Section~\ref{excit} has shown that optimal currents obtained by the method described in this paper should be interpreted mostly as unbreakable and unreachable lower or upper bounds. Any attempt to create realistic feeding of such current will always lead to a sub-optimal solution. An interesting possibility of approaching the optimal current is to utilize characteristic modes (\ref{Qopt01}) and try to excite them selectively by a combination of feeders and slots in a metallic support covering the optimized region \cite{MartensManteuffel_SystematicDesignMethodOfMobileMultipleAntennaSystemUsingCM}. The optimization scheme presented in this paper gives the optimal modal excitation coefficients.


\section{Conclusion}
\label{concl}
A systematic approach leading to a surface current density optimizing convex and non-convex radiation metrics has been presented. The method has been used to find a current density minimizing quality factor $Q$ and ohmic losses which are two core parameters of electrically small antennas. The proposed method is, however, of general validity not being limited to an electrically small domain.

The optimization has been presented in the RWG basis and it has been demonstrated that forming the optimal current as a combination of modes of specific generalized eigenvalue problem can tremendously simplify the solution. In many relevant cases the optimum can be reached by the proper combination of only few modes. The form of the eigenvalue problem has been specified for the optimization of quality factor $Q$, gain $G$, the $G/Q$ ratio and radiation efficiency $\eta$. 
 
The presented method provides results fully consistent with previous works, but offers the added advantage of producing upper and lower bounds of radiation parameters for surfaces of arbitrary shape. Of particular importance is the fact that the method enables us to find the optimal combination of characteristic modes for specified optimized parameter. This can lead to the realistic excitation of the optimal current density and, thus, to the design of optimal antennas.

Future work should aim at finding a realistic excitation of the optimal current densities as well as a way to utilize the method for electrically large structures, such as antenna arrays.

\section*{Acknowledgement}
The authors would like to thank Mats Gustafsson from the University of Lund (Sweden) for his thoughts and suggestions which stimulated the development of the presented work.

\section{Appendix}
\label{app}
This appendix contains explicit forms of matrices $\mathbf{R}$, $\mathbf{X}$, $\mathbf{W}$, $\mathbf{\Sigma}$, $\mathbf{U}_{\left( {\theta /\varphi} \right)}$ in the RWG representation which assumes the expansion of the surface current density into the set of RWG functions as \cite{RaoWiltonGlisson_ElectromagneticScatteringBySurfacesOfArbitraryShape} 
\BE
\label{appA1}
{\boldsymbol{J}}\left( {\boldsymbol{r}} \right) = \sum\limits_n {{I_n}{{\boldsymbol{f}}_n}\left( {\boldsymbol{r}} \right)},
\EE
where $I_n$ are the RWG edge surface current densities \cite{RaoWiltonGlisson_ElectromagneticScatteringBySurfacesOfArbitraryShape}.

\subsection{Radiated and reactive power matrices}
\label{appRX}
Within the notation used, the complex power \cite{Harrington_TimeHarmonicElmagField} can be written as
\BE
\label{AppA01}
{P_{{\mathrm{rad}}}} + {\mathrm{j}}{P_{{\mathrm{react}}}} = \frac{{\mathrm{j}}}{{8{\mathrm{\pi }}{\varepsilon _0}\omega }}\left( {{k^2}\left\langle {{\boldsymbol{J}},L{\boldsymbol{J}}} \right\rangle  - \left\langle {\nabla  \cdot {\boldsymbol{J}},L\nabla  \cdot {\boldsymbol{J}}} \right\rangle } \right).
\EE
Substitution of (\ref{appA1}) directly leads to real symmetric matrices $\mathbf{R}$, $\mathbf{X}$ in the form
\BE
\label{AppA02}
\mathbf{R} + {\mathrm{j}}\mathbf{X} = {\mathrm{j}}\left[ {\frac{1}{{8{\mathrm{\pi }}{\varepsilon _0}\omega }}\left( {{k^2}\left\langle {{{\boldsymbol{f}}_m},L{{\boldsymbol{f}}_n}} \right\rangle  - \left\langle {\nabla  \cdot {{\boldsymbol{f}}_m},L\nabla  \cdot {{\boldsymbol{f}}_n}} \right\rangle } \right)} \right],
\EE
where operator $L$ is defined by (\ref{eqS14a}). Consulting the form of (\ref{AppA02}) with \cite{RaoWiltonGlisson_ElectromagneticScatteringBySurfacesOfArbitraryShape} reveals that \mbox{$\mathbf{R} + {\mathrm{j}}\mathbf{X}$} is just the EFIE impedance matrix \cite{Harrington_FieldComputationByMoM,Gustafsson_OptimalAntennaCurrentsForQsuperdirectivityAndRP}.

\subsection{Stored energy matrix}
\label{appWsto}
According to \cite{Vandenbosch_ReactiveEnergiesImpedanceAndQFactorOfRadiatingStructures}, stored electromagnetic energy can be evaluated as
\BE
\label{AppC01}
\begin{aligned}
	{W_{{\mathrm{sto}}}} &= \frac{1}{{{\mathrm{16\pi }}{\varepsilon _0}{\omega ^2}}}{\mathrm{Re}}  \Big \{ {k^2}\left\langle {{\boldsymbol{J}},L{\boldsymbol{J}}} \right\rangle  + \left\langle {\nabla  \cdot {\boldsymbol{J}},L\nabla  \cdot {\boldsymbol{J}}} \right\rangle \\
	&- {\mathrm{j}}k\left( {{k^2}\left\langle {{\boldsymbol{J}},{L_{{\mathrm{rad}}}}{\boldsymbol{J}}} \right\rangle  - \left\langle {\nabla  \cdot {\boldsymbol{J}},{L_{{\mathrm{rad}}}}\nabla  \cdot {\boldsymbol{J}}} \right\rangle } \right) \Big \}
\end{aligned}
\EE
with
\BE
\label{AppC01a}
	{L_{{\mathrm{rad}}}}{\boldsymbol{J}} = \int\limits_{V'} {{\boldsymbol{J}}\left( {{\boldsymbol{r'}}} \right){{\mathrm{e}}^{ - {\mathrm{j}}k\left| {{\boldsymbol{r}} - {\boldsymbol{r'}}} \right|}}{\mathrm{d}}V'}.
\EE
Substituting (\ref{appA1}) into (\ref{AppC01}) leads to the real symmetric matrix
\BE
\label{AppC02}
\begin{aligned}
\mathbf{W} &= \frac{1}{{16{\mathrm{\pi }}{\varepsilon _0}{\omega ^2}}} \mathrm{Re} \Big \{ \Big[ {k^2}\left\langle {{{\boldsymbol{f}}_m},L{{\boldsymbol{f}}_n}} \right\rangle  + \left\langle {\nabla  \cdot {{\boldsymbol{f}}_m},L\nabla  \cdot {{\boldsymbol{f}}_n}} \right\rangle  \\
&- {\mathrm{j}}k\left( {{k^2}\left\langle {{{\boldsymbol{f}}_m},{L_{{\mathrm{rad}}}}{{\boldsymbol{f}}_n}} \right\rangle  - \left\langle {\nabla  \cdot {{\boldsymbol{f}}_m},{L_{{\mathrm{rad}}}}\nabla  \cdot {{\boldsymbol{f}}_n}} \right\rangle } \right) \Big] \Big \}.
\end{aligned}
\EE
Following the reasoning in \cite{Gustafsson_OptimalAntennaCurrentsForQsuperdirectivityAndRP,Gustaffson_QdisperssiveMedia_arXiv,GustafssonFridenColombi_AntennaCurrentOptimizationForLossyMediaAWPL} it can be shown that
\BE
\label{AppC03}
\mathbf{W} = \frac{1}{4}\frac{{\partial {\mathbf{X}}}}{{\partial \omega }}.
\EE

\subsection{Lost power matrix}
\label{appSigma}
The cycle mean power lost at the surface of a good conductor can be written as \cite{Jackson_ClassicalElectrodynamics}
\BE
\label{AppD01}
{P_{{\mathrm{lost}}}} = \frac{1}{2}\left\langle {{\boldsymbol{J}},\mathrm{Re} \left\{ {{Z_\mathrm{s}}} \right\}{\boldsymbol{J}}} \right\rangle,
\EE
where \mbox{${Z_{\mathrm{s}}} = \left( {1 + {\mathrm{j}}} \right)/\left( {\sigma \delta } \right)$} is the surface impedance of the conducting half-space \cite{Jackson_ClassicalElectrodynamics} and \mbox{$\delta  = \sqrt {2/\left( {\omega \mu \sigma } \right)}$} is the penetration depth. The symmetric and real matrix representing losses directly follows in the form
\BE
\label{AppD02}
\mathbf{\Sigma} = \mathrm{Re} \left\{ {{Z_\mathrm{s}}} \right\} \Big[ {\left\langle {{{\boldsymbol{f}}_m},{{\boldsymbol{f}}_n}} \right\rangle } \Big],
\EE
where an assumption of the surface impedance ${{Z_\mathrm{s}}}$ being constant within the range of (\ref{AppD01}) has been used. A straightforward integration in barycentric coordinates gives
\BE
\label{AppD03}
\begin{aligned}
\left\langle {{{\boldsymbol{f}}_m},{{\boldsymbol{f}}_m}} \right\rangle  &= \frac{{l_m^2}}{{24A_m^ + }}\Bigg[ {\boldsymbol{r}}_m^{\left( {{\mathrm{c}+}} \right)} \cdot \left( {9{\boldsymbol{r}}_m^{\left( {{\mathrm{c}+}} \right)} - 15{\boldsymbol{p}}_m^{\left( 1 \right)}} \right) \\
&+ 7{{\left| {{\boldsymbol{p}}_m^{\left( 1 \right)}} \right|}^2} - {\boldsymbol{p}}_m^{\left( 2 \right)} \cdot {\boldsymbol{p}}_m^{\left( 3 \right)} \Bigg] + \\
&+ \frac{{l_m^2}}{{24A_m^ - }}\Bigg[ {\boldsymbol{r}}_m^{\left( {{\mathrm{c}}-} \right)} \cdot \left( {9{\boldsymbol{r}}_m^{\left( {{\mathrm{c}}-} \right)} - 15{\boldsymbol{p}}_m^{\left( 4 \right)}} \right) \\
&+ 7{{\left| {{\boldsymbol{p}}_m^{\left( 4 \right)}} \right|}^2} - {\boldsymbol{p}}_m^{\left( 2 \right)} \cdot {\boldsymbol{p}}_m^{\left( 3 \right)} \Bigg]
\end{aligned}
\EE
for diagonal terms and
\BE
\label{AppD04}
\begin{aligned}
\left\langle {{{\boldsymbol{f}}_m},{{\boldsymbol{f}}_n}} \right\rangle  &= \frac{{{\chi _{mn}}{l_m}{l_n}}}{{24{A_m}}}\Bigg[ 9{\boldsymbol{r}}_m^{\left( {\mathrm{c}} \right)} \cdot \left( {{\boldsymbol{r}}_m^{\left( {\mathrm{c}} \right)} - {\boldsymbol{p}}_m^{\left( {\mathrm{f}} \right)} - {\boldsymbol{p}}_n^{\left( {\mathrm{f}} \right)}} \right) \\
&+ {{\left| {{\boldsymbol{p}}_m^{\left( {\mathrm{f}} \right)} + {\boldsymbol{p}}_n^{\left( {\mathrm{f}} \right)}} \right|}^2} + 5{\boldsymbol{p}}_m^{\left( {\mathrm{f}} \right)} \cdot {\boldsymbol{p}}_n^{\left( {\mathrm{f}} \right)} \Bigg]
\end{aligned}
\EE
for off-diagonal terms, with $l_m$ as the edge length of the \mbox{$m$-th} RWG function, ${A_m^{\pm} }$ as the area of its positive / negative triangle and ${\boldsymbol{r}}_m^{\left({\mathrm{c}} \pm \right)}$ as the positive / negative triangle centre
\cite{RaoWiltonGlisson_ElectromagneticScatteringBySurfacesOfArbitraryShape}. The vertices $\boldsymbol{p}$ are defined according to Fig.~\ref{fig14}a. The \mbox{superindex $^{\left( {\mathrm{f}} \right)}$}, used in (\ref{AppD04}), denotes free vertices (the vertices $\boldsymbol{p}^{\left( 1 \right)}$ and $\boldsymbol{p}^{\left( 4 \right)}$) belonging to the triangle common to the \mbox{$m$-th} and the \mbox{$n$-th} RWG function. The coefficient $\chi_{mn}$ is equal to unity for cases depicted in Fig.~\ref{fig14}b,c, to minus unity for cases depicted in Fig.~\ref{fig14}d,e and to zero for RWG functions with no common triangle.
\BF
\begin{center}
	\includegraphics[width=8.1cm]{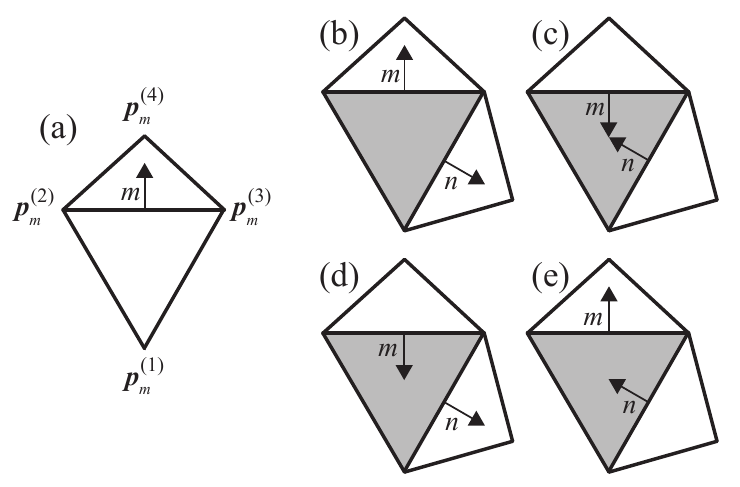}
	\caption{Sketch of the $m$-th RWG function (a) and of an overlap (b, c, d, e) between the $m$-th and the $n$-th RWG function. The orientation of the RWG function is denoted by an arrow. The vertices are denoted by corresponding radius vectors $\boldsymbol{p}_m$. The grey colour represents the overlap region.}
	\label{fig14}
\end{center}
\EF

When the current layer is not thick enough with respect to the penetration depth, the aforementioned formulation can be significantly improved by changing \cite{CapekEichlerHazdra_EvaluationOfRadiationEfficienciIET}
\BE
\label{AppD03}
\displaystyle \mathrm{Re} \left\{ {{Z_{\mathrm{s}}}} \right\} \to \frac{1}{{\sigma \delta }}\left( {1 - {{\mathrm{e}}^{ - 2\frac{t}{\delta }}}} \right){\left| {1 - {{\mathrm{e}}^{ - \left( {1 - {\mathrm{j}}} \right)\frac{t}{\delta }}}} \right|^{ - 2}},
\EE
where $t$ is the thickness of the lossy layer.

\subsection{Radiation intensity matrix}
\label{appU}
The partial radiation intensity $U_{\left( {\theta /\varphi } \right)}$ in spherical direction $\boldsymbol{\theta}_0$ or $\boldsymbol{\varphi}_0$ is related to partial directivity \cite{Balanis_Wiley_2005} as
\BE
\label{appB1}
{D_{\left( {\theta /\varphi } \right)}} = \frac{{4\pi {U_{\left( {\theta /\varphi } \right)}}}}{{{P_{{\mathrm{rad}}}}}} = \frac{{2\pi {{\left| {{F_{\left( {\theta /\varphi } \right)}}} \right|}^2}}}{{{Z_0}{P_{{\mathrm{rad}}}}}},
\EE
where ${P_{{\mathrm{rad}}}}$ is the radiated power (\ref{AppA01}) and
\BE
\label{appB2}
{F_{\left( {\theta /\varphi } \right)}} = \frac{{ - {\mathrm{j}}\omega \mu }}{{4\pi }}\int\limits_{S'} {{J_{\left( {\theta /\varphi } \right)}}\left( {{\boldsymbol{r'}}} \right){{\mathrm{e}}^{{\mathrm{j}}k{{\boldsymbol{r}}_0} \cdot {\boldsymbol{r'}}}}{\mathrm{d}}S'}
\EE
is the far-field radiation pattern projected in a given spherical direction \cite{Balanis_Wiley_2005}. Note that the observation directions $\boldsymbol{\theta}_0$, $\boldsymbol{\varphi}_0$, $\boldsymbol{r}_0$ are assumed constant during the integration (\ref{appB2}).

The partial radiation intensity can, therefore, be written as
\BE
\label{appB3}
{U_{\left( {\theta /\varphi } \right)}} = \frac{{{Z_0}{k^2}}}{{32{\pi ^2}}}\left\langle {{J_{\left( {\theta /\varphi } \right)}},{L_U}{J_{\left( {\theta /\varphi } \right)}}} \right\rangle	
\EE
with
\BE
\label{appB4}
{L_U}M = \int\limits_{S'} {M\left( {{\boldsymbol{r}'}} \right){{\mathrm{e}}^{ - {\mathrm{j}}k{{\boldsymbol{r}}_0} \cdot \left( {{\boldsymbol{r}} - {\boldsymbol{r}'}} \right)}}{\mathrm{d}}S'}.
\EE	
Assuming the expansion (\ref{appA1}), the partial radiation intensity matrix ${\mathbf{U}_{\left( {\theta /\varphi } \right)}}$ can be written as
\BE
\label{appB7}
{\mathbf{U}_{\left( {\theta /\varphi } \right)}} = \frac{{{Z_0}{k^2}}}{{32{\pi ^2}}}\Big[ \left\langle {{f_{\left( {\theta /\varphi } \right)m}},{L_U}{f_{\left( {\theta /\varphi } \right)n}}} \right\rangle \Big].
\EE
Assuming futher that RWG triangles are much smaller than the operating wavelength, we can approximate
\BE
\label{appB8}
\int\limits_{S'} {{f _{\left( {\theta /\varphi } \right)n}}\left( {\boldsymbol{r}'} \right){{\mathrm{e}}^{{\mathrm{j}}k{{\boldsymbol{r}}_0} \cdot {\boldsymbol{r}'}}}{\mathrm{d}}S'}  \approx {{\mathrm{e}}^{{\mathrm{j}}k{{\boldsymbol{r}}_0} \cdot \frac{{{\boldsymbol{r}}_n^{\left(\mathrm{c}+\right)} + {\boldsymbol{r}}_n^{\left(\mathrm{c}-\right)}}}{2}}}\int\limits_S {{f _{\left( {\theta /\varphi } \right)n}}{\mathrm{d}}S}.
\EE
This allows for writing
\BE
\label{appB8}
{{\mathbf{U}}_{\left( {\theta /\varphi } \right)}} = \frac{{{Z_0 k^2}}}{{32{\pi ^2}}}{\mathbf{u}}_{\left( {\theta /\varphi } \right)}^{\mathrm{H}}{{\mathbf{u}}_{\left( {\theta /\varphi } \right)}},
\EE
where ${{\mathbf{u}}_{\left( {\theta /\varphi } \right)}}$ is a column vector of components
\BE
\label{appB9}
{u_{\left( {\theta /\varphi } \right)n}} = {l_n}\left( {r_{\left( {\theta /\varphi } \right)n}^{\left(\mathrm{c}+\right)} - r_{\left( {\theta /\varphi } \right)n}^{\left(\mathrm{c}-\right)}} \right){{\mathrm{e}}^{{\mathrm{j}}k{{\mathbf{r}}_0} \cdot \frac{{{\mathbf{r}}_n^{\left(\mathrm{c}+\right)} + {\mathbf{r}}_n^{\left(\mathrm{c}-\right)}}}{2}}}.
\EE

\ifCLASSOPTIONcaptionsoff
  \newpage
\fi

\bibliographystyle{IEEEtran}

\begin{biography}[{\includegraphics[width=1in,height=1.25in,clip,keepaspectratio]{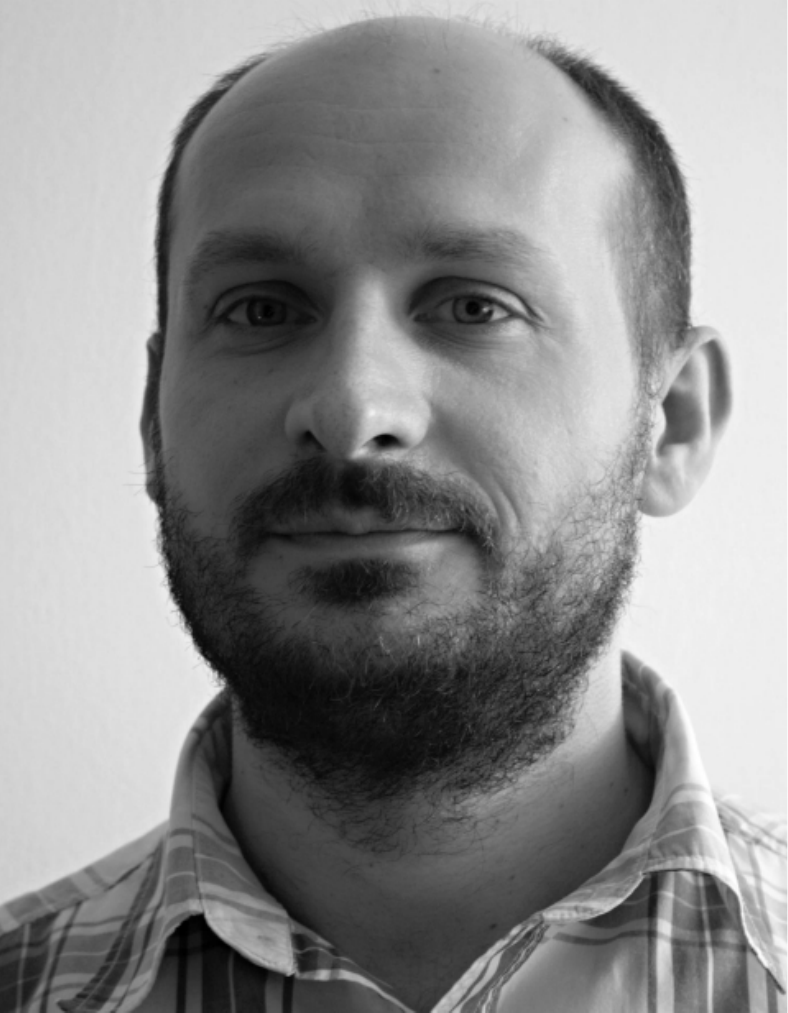}}]{Lukas Jelinek}
received his Ph.D. degree from the Czech Technical University in Prague, Czech Republic, in 2006. In 2015 he was appointed Associate Professor at the Department of Electromagnetic Field at the same university.

His research interests include wave propagation in complex media, general field theory, numerical techniques and optimization.
\end{biography}
\begin{biography}[{\includegraphics[width=1in,height=1.25in,clip,keepaspectratio]{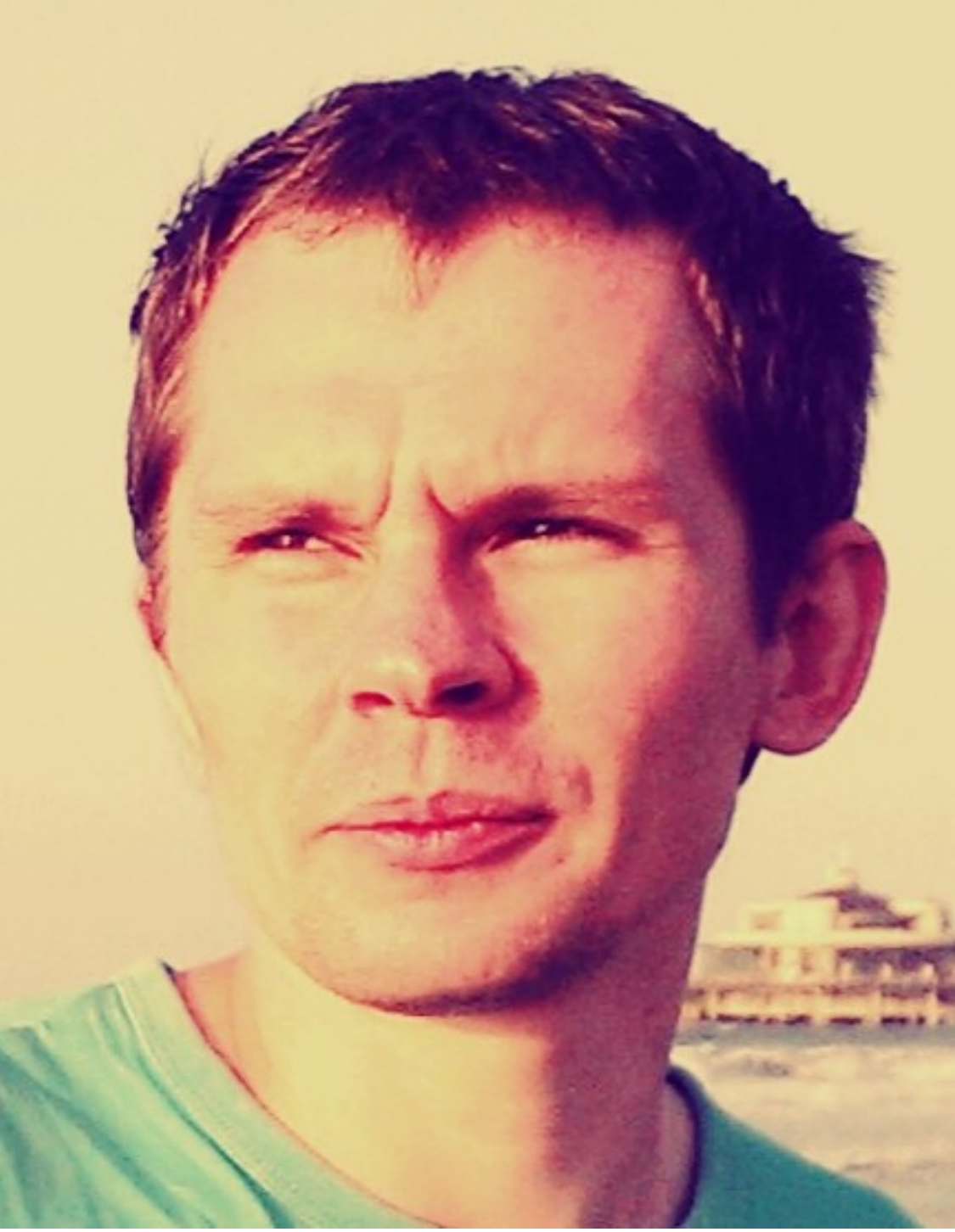}}]{Miloslav Capek}
	(S'09, M'14) received his M.Sc. degree in Electrical Engineering from the Czech Technical University, Czech Republic, in 2009, and his Ph.D. degree from the same University, in 2014. Currently, he is a researcher with the Department of Electromagnetic Field, CTU-FEE.
	
	He leads the development of the AToM (Antenna Toolbox for Matlab) package. His research interests are in the area of electromagnetic theory, electrically small antennas, numerical techniques, fractal geometry and optimization. He authored or co-authored over 45 journal and conference papers.
	
	Dr. Capek is member of Radioengineering Society, regional delegate of EurAAP, and Associate Editor of Radioengineering.
\end{biography}

\end{document}